\documentclass{article}

\usepackage{amsmath}
\usepackage{amsfonts}
\usepackage{amssymb}
\usepackage{amscd}
\usepackage{amsthm}

\def\R{\mathbb{R}}
\def\C{\mathbb{C}}
\def\T{\mathbb{T}}
\def\Z{\mathbb{Z}}
\def\N{\mathbb{N}}
\def\H{\mathcal H}
\def\K{\mathcal K}
\def\Op{\mathfrak{Op}}
\def\End{\mathfrak{End}}
\def\Aut{\mathfrak{Aut}}
\def\Rep{\mathfrak{Rep}}
\def\M{\mathfrak{M}}
\def\U{\mathcal U}
\def\F{\mathcal F}
\def\S{\mathcal S}
\def\A{{\mathcal A}}
\def\I{{\mathcal I}}
\def\B{{\mathcal B}}
\def\ST{s_{\hbox{\tiny T}}}
\def\SA{S_{\!\A}}

\def\e{\varepsilon}
\def\1{\mathfrak{1}}
\def\0{\mathfrak{0}}
\def\G{\mathfrak{G}}
\def\n{|\!|\!|}
\def\d{\diamond}

\def\CO{{\mathcal C}_0}
\def\BC{\mathcal{BC}}
\def\CC{{\mathcal C}}
\def\FX{{\mathcal F}_{\!\hbox{\tiny \it X}}}
\def\FXb{\overline{{\mathcal F}}_{\!\hbox{\tiny \it X}}}
\def\FXS{{\mathcal F}_{\!\hbox{\tiny \it X}^\sharp}}
\def\FXSb{\overline{{\mathcal F}}_{\!\hbox{\tiny \it X}^\sharp}}
\def\FRb{\overline{{\mathcal F}}_{\!\hbox{\tiny \it X}^\star}}

\newtheorem{lemma}{Lemma}[section]

\newtheorem{proposition}[lemma]{Proposition}
\newtheorem{definition}[lemma]{Definition}
\newtheorem{remark}[lemma]{Remark}

\numberwithin{equation}{section}

\begin{document}

\title{Twisted Crossed Products \\ and Magnetic Pseudodifferential
Operators}

\date{\today}

\author{Marius M\u antoiu, Radu Purice\footnote{Institute
of Mathematics ``Simion Stoilow'' of
the Romanian Academy, P.O.  Box 1-764, Bucharest, RO-70700, Romania,
Email: mantoiu@imar.ro, purice@imar.ro}
\ and Serge Richard\footnote{Department of Theoretical Physics,
University of Geneva, 1211 Geneva 4, Switzerland,
Email: serge.richard@physics.unige.ch}}

\maketitle

\begin{abstract}
There is a connection between the Weyl pseudodifferential calculus
and crossed product $C^*$-algebras associated with certain dynamical
systems. And in fact both topics are involved in the quantization of
a non-relativistic particle moving in $\R^N$. Our paper studies the
situation in which a variable magnetic field is also present.
The Weyl calculus has to be modified, giving a functional calculus for
a family of operators (positions and magnetic momenta) with highly
non-trivial commutation relations. On the algebraic side, the
dynamical system is twisted by a cocycle defined by the flux of the
magnetic field, leading thus to twisted crossed products. Following
mainly \cite{Purice1} and \cite{Purice2}, we outline the interplay
between the modified pseudodifferential setting and the
$C^*$-algebraic formalism at an abstract level as well as in
connection with magnetic fields.
\end{abstract}



\tableofcontents

\section*{Introduction}

One of the  purposes of the present article is to study the
mathematical objects involved in the quantization of physical
systems placed in a magnetic field. Two structures emerge
naturally and in a correlated manner: (1) a modified form
of the pseudodifferential Weyl calculus and (2) twisted
crossed product $C^*$-algebras. In fact, the connection
between (1) and (2) is present in a more general setting and
showing this is also one of our aims. Under favorable circumstances,
both (1) and (2) involve algebras of symbols defined on variables
having a physical interpretation and the representations of these
algebras may be seen as functional calculi associated with families
 of non-commuting observables.

Let us first explain roughly what we mean by ``quantization''.
We assume that a given physical system admits {\it a phase space},
at least in a weak sense. We do not give a precise definition of
this concept, but approximatively it is a space containing the
canonical coordinates, often interpreted as ``positions'' and
``momenta". These coordinates enter in the equations that describe
in Classical Mechanics the time evolution of the system.
The phase space is usually modelled mathematically by a symplectic
manifold and its points are called {\it the states of the system}.
The {\it observables}, the quantities that can be measured, are
described by (smooth) functions on the phase space. On the other
hand, the quantum description of the same system requires a Hilbert
space and the observables are represented by self-adjoint or normal
operators acting in the Hilbert space. Then a {\it quantization}
would be a systematic procedure to assign quantum observables
(operators) to classical ones (functions). This procedure should obey
a set of principles, which are difficult to state once for all in a
universal way.

We stress that - for the moment - our approach is only a first
(important) step towards a deformation quantization \`a la Rieffel
(cf.~\cite{Rieffel3} or \cite{Landsman}). All is done for a fixed value of
Planck's constant; in fact we take $\hbar=1$. We hope to be able to let
$\hbar$ vary in a subsequent publication.

We also note that the word ``phase space'' should not be taken too
strictly; in particular it can differ from a cotangent bundle. If the
configuration space is taken to be $\Z^N$ for instance, there
is no cotangent bundle, but aside multiplication operators, one is
also interested in finite-difference operators. Actually our ``phase
space'' $\Xi$ will be the direct product of an abelian, locally compact
group $X$ and its dual $X^\sharp$. $X$ serves as a configuration space
and $X^\sharp$ is somehow connected with a momentum observable, but
this may be rather vague if not enough structure is present. Anyhow,
the main interest of our work lies in the fact that this momentum
observable can be ``twisted'', the main example being the magnetic
momentum. This means that it may correspond not to a usual action
$\theta$ in the configuration space, but to a twisted action
$(\theta,\omega)$, where $\omega$ is a $2$-cocycle of the group $X$
with values in some general functions space.

Let us comment on the most important case, $X=\R^N$. A common point
of view is that the Weyl form of the usual pseudodifferential calculus
in $\R^N$ (cf.~\cite{Folland}, \cite{Hormander}, \cite{Shubin1}) is a
functional calculus for the family of operators
$(Q_1,\dots,Q_N,P_1,\dots,P_N)$, where $Q_j$ is the operator of
multiplication by the $j$'th coordinate and $P_j=-i\partial_j$. This
calculus is suited to the description of a non-relativistic quantum
particle moving in $\R^N$ and having no internal structure. Its
complexity is due to the fact that the operators $Q_j$ and $P_j$ do not
commute. If the particle is placed in a magnetic field $B$, the
momentum $P_j$ has to be replaced by the magnetic momentum
$\Pi^A_j=P_j-A_j(Q)$, where $A$ is a vector potential assigned to the
magnetic field: $B=dA$. The problem of constructing a functional
calculus for the more complicated family
$(Q_1,\dots,Q_N,\Pi^A_1,\dots,\Pi^A_N)$ was tackled in \cite{Karasev1},
\cite{Karasev2} and \cite{Purice2}; the result was called {\it the magnetic
Weyl calculus}. Now there is a higher (but still manageable) degree of
non-commutativity, in terms of phase factors defined by the flux of the
magnetic field through suitable triangles and the circulation of the
vector potential through suitable segments.

Crossed product $C^*$-algebras (for which we refer to \cite{Pedersen}
or to \cite{Williams}) originated in physics (cf.~\cite{Doplicher})
and had an exceptional career, very often independent of the initial
motivation. There also exists a more involved version, but still very
natural, called {\it twisted crossed product}. It was initiated in
\cite{Busby} and found its full strength in \cite{Raeburn1} and
\cite{Raeburn2}; a useful review is \cite{Packer}. The motivations
were of a pure mathematical nature. It was shown
in \cite{Purice1} that the concept of twisted crossed product
$C^*$-algebra is extremely natural in connection with the quantization
of physical systems in the presence of magnetic fields.
The point is that a reformulation of the magnetic commutation
relations puts into evidence {\it a twisted dynamical system},
the twist being related to the flux of the magnetic field.
This reveals automatically a formalism which is in a certain sense
isomorphic to the magnetic Weyl calculus. But the $C^*$-algebraic
version is more precise and flexible. It also gives a more transparent
view on cohomological matters related to gauge covariance.

Essentially, this is an expository article, summarizing results of
\cite{Purice1} and \cite{Purice2} and underlining the connection
between them. But a large part of the approach is rephrased and
several new results are included.

Let us describe the content. First of all, in an introductory section, we
review briefly some basic facts about the quantization of a non-relativistic,
spinless particle moving in the flat $N$-dimensional configuration
space (in the absence of any magnetic field). This should be
a motivation for the main body of the text, since the key notions
(canonical commutation relations, Weyl calculus, dynamical systems and
their crossed products) appear already in a simple form.

In Section \ref{s2} we review twisted dynamical systems, covariant
representations and twisted crossed products. We are placed in a
general setting, but with some simplifying assumptions which should
make this topic more popular to non-specialists.  Some general facts
in group cohomology are also presented briefly, leading to a general
form of gauge covariance. The assumption that the algebra on which the
group acts is actually composed of bounded, uniformly continuous
functions defined on the group leads to several specific properties,
as the existence of a general class of Schr\"odinger-type representations.
This also eases the way to the next section.

In Section \ref{s3}, by a simple reformulation, we get objects
generalizing in some sense the pseudodifferential calculus. It might be
instructive to note that parts of the pseudodifferential theory
depend only on a rather general setting and still work for
commutation relations more involved than those satisfied by $Q$ and $P$.

In Section \ref{s4} we show how all these particularize to the twisted
dynamical system associated with a particle in a magnetic field. This
physical case is the main motivation of our work.

Our interest in twisted crossed products and in the magnetic Weyl
calculus lies beyond this general setting. In a forthcoming
publication we will show that  the twisted crossed product $C^*$-algebras
contain the functional calculus of suitable classes of
Schr\"odinger operators with magnetic fields. This opens the way
towards extending spectral or propagation results as
those of \cite{Amrein}, \cite{Georgescu1}, \cite{Mantoiu1},
\cite{Amrein'} to magnetic Schr\"odinger operators.
This will be done by $C^*$-algebraic techniques, involving
constructions and results of the present paper.
 
Our hope is that our approach could be interesting for at least three
groups of people: $C^*$-algebraists (especially those involved in
$C^*$-dynamical systems), people working in pseudodifferential theory and
mathematical physicists interested in quantum systems with magnetic fields.
Since none of the three needs to be an expert in the other two topics,
we shall try to be rather elementary. We also defer more
deep or more technical developments as well as applications to future
works. The present purpose is just to show that a new
pseudodifferential calculus is well justified physically, has a life of
its own and can be recast in the language of twisted crossed products.
A really critical point is $C^*$-algebras. Although they are no longer
considered to be a close friend of the mathematical physicists working
in quantum theory, our opinion is that they provide very useful techniques
and insights and we intend to continue to use them in connection with
magnetic fields. If one is interested in pseudodifferential operators with
magnetic fields but without $C^*$-algebras,
(s)he could consult \cite{Karasev1}, \cite{Karasev2} and \cite{Purice2}.

\section{Zero magnetic field - a heuristic presentation}

The class of physical systems for which a reliable quantization is
already achieved is rather poor. The outstanding example is certainly
the non-relativistic quantum particle moving in $\R^N$ and having no
additional (internal) degree of freedom. We describe briefly this
situation before introducing any magnetic field, since this serves
both as a motivation and as a comparison theory.

\subsection{The Weyl calculus - a paradigm of quantization}\label{sub11}

We consider a non-relativistic particle without internal structure,
moving in the configuration space $X \equiv \R^N$. The phase space
of this system is the cotangent bundle $T^*X = X\times X^\star\equiv
\R^{2N}$, denoted from now on by $\Xi$, on which we have the canonical
symplectic form
\begin{equation*}
\sigma(\xi,\eta)=\sigma\left[(x,p),(y,k)\right]:=y \cdot p - x \cdot k,
\end{equation*}
where $y \cdot p$ is the action of the linear functional $p\in X^\star
\equiv \R^N$ on the vector $y\in X$. Thus a classical observable is
just a (smooth real or complex) function defined on $T^*X$. The
symplectic form serves among others in defining the Poisson bracket
\begin{equation*}
\{f,g\}:=\sigma(\nabla f,\nabla g)=\sum_{j=1}^N
\left(\partial_{p_j}f\;\partial_{q_j}g-
\partial_{p_j}g\ \partial_{q_j}f\right),
\end{equation*}
endowing the set of observables with the structure of a Lie algebra.
Here we have made the usual abuse of identifying the tangent space
at a point $\xi\in\Xi$ with the linear space $\Xi$ itself and thus
extending the canonical symplectic form $\sigma$ to the tangent bundle
$T\Xi$. This Poisson bracket plays an important role in the Hamiltonian
formulation of the equations of motion.

Let us now describe briefly and rather formally the quantization of
this system. One first deals with the coordinate functions $q_1,\dots
q_N$ and $p_1,\dots,p_N$ satisfying the relations
\begin{equation*}
\{q_i,q_j\}=0,\ \{p_i,p_j\}=0,\ \{p_i,q_j\}=\delta_{ij}, \qquad
i,j=1,\dots,N.
\end{equation*}
A dogma of the physical community is to assign to them the self-adjoint
operators $Q_1,\dots,Q_N,P_1,\dots,P_N$ satisfying the same relations
but with the Poisson bracket $\{\cdot,\cdot\}$ replaced by
$i[\cdot,\cdot]$ (here $[S,T]:=ST-TS$ is the commutator).
Thus we should have
\begin{equation}\label{PQ}
i[Q_i,Q_j]=0,\ i[P_i,P_j]=0,\ i[P_i,Q_j]=\delta_{ij}, \qquad i,j=1,
\dots,N.
\end{equation}
It is known that if one also asks the irreducibility of the family
$Q_1,\dots,Q_N,P_1,\dots,P_N$ and identifies unitarily
equivalent families, then there is only one possible choice for $Q$
and $P$\;: the Hilbert space is $\H := L^2(X)$, $Q_j$ is the operator
of multiplication by $x_j$ and $P_j$ is the differential operator
$-i\partial_j$. For the exact form of this statement,
involving the unitary groups generated by $Q$ and $P$, called {\it
the Stone-von Neumann Theorem}, see \cite{Folland}.
 
The natural next step is the quantization of more general functions.
This can be seen as the problem of constructing a functional calculus
$f\mapsto f(Q,P)$ for the family $Q_1,\dots,Q_N, P_1,\dots,P_N$ of
$2N$ self-adjoint, non-commuting operators. One would also like to define
a ``quantum'' multiplication $(f,g)\mapsto f\circ g$ satisfying $(f\circ g)
(Q,P)=f(Q,P)g(Q,P)$ as well as an involution $f\to f^{\circ}$ leading
to $f^\circ(Q,P)=f(Q,P)^*$. The deviation of $\circ$ from pointwise
multiplication is imputable to the fact that $Q$ and $P$ do not commute.
The solution of these problems is called {\it the Weyl calculus}.
The prescription is $f(Q,P)=\Op(f)$, with
\begin{equation}\label{Weyl}
[\Op(f)u](x):=\int_{\R^{2N}}dy\;\!dp\;e^{i(x-y)\cdot
p}f\left(\frac{x+y}{2},p\right)u(y), \qquad u\in \H,
\end{equation}
the involution is $f^\circ(\xi):=\overline{f(\xi)}$ and the
multiplication (called {\it the Moyal product}) is
\begin{equation}\label{Moyal}
(f\circ g)(\xi):=4^N\int_\Xi d\eta\int_\Xi d\zeta\;e^{-2i\sigma(\xi
-\eta,\xi-\zeta)}f(\eta)\;\!g(\zeta), \qquad \xi\in\Xi.
\end{equation}
The two formulae must be taken with some care: for many symbols
$f$ and $g$ they need a suitable reinterpretation.

Let us try to show where all these come from.
We consider the strongly continuous unitary maps $X \ni x
\mapsto U(x):=e^{-ix\cdot P} \in \U(\H)$ and $X^\star \ni p
\mapsto V(p):=e^{-iQ\cdot p} \in \U(\H)$, acting on $\H$ as
\begin{equation*}
[U(x)u](y)=u(y-x) \quad \text{and} \quad [V(p)u](y)=e^{-iy\cdot p}
\;u(y), \qquad u \in \H,\ y\in X.
\end{equation*}
These operators satisfy the {\it Weyl form of the canonical
commutation relations}
\begin{equation}\label{WCCR}
U(x)V(p)=e^{ix\cdot p}\;V(p)U(x),\qquad x\in X,\ p\in X^\star,
\end{equation}
as well as the identities $U(x)U(x')=U(x')U(x)$ and
$V(p)V(p')=V(p')V(p)$ for $x,x' \in X$ and $p,p' \in X^\star$. These
can be considered as a reformulation of (\ref{PQ}) in terms of bounded
operators.

A convenient way to condense the maps $U$ and $V$ in a single one
is to define {\it the Schr\"odinger Weyl system} $\{W(x,p)\;|\;x\in X,
\ p \in X^\star\}$ by
\begin{equation}\label{wey}
W(x,p):=e^{\frac{i}{2}x\cdot p}\;U(-x)V(p)=e^{-\frac{i}{2}x\cdot p}
\;V(p)U(-x),
\end{equation}
which satisfies the relation $W(\xi)W(\eta)=e^{\frac{i}{2}
\sigma(\xi,\eta)}\;W(\xi+\eta)$ for any $\xi,\eta\in\Xi$.
This equality encodes all the commutation relations between the basic
operators $Q$ and $P$. Explicitly, the action of $W$ on $u \in \H$
is given by
\begin{equation}\label{Waction}
[W(x,p)u](y)=e^{-i\left(\frac{1}{2}x+y\right)\cdot p}\;u(y+x),
\qquad x,y \in X,\ p \in X^\star.
\end{equation}

For a family of $m$ commuting self-adjoint operators $S_1, \ldots,
S_m$ one usually defines a functional calculus by the formula
$f(S):=\int_{\R^m}dt\;\check{f}(t)e^{-it\cdot S}$, where $t \cdot S =
t_1 S_1 + \ldots + t_m S_m$ and $\check{f}$ is the inverse Fourier
transform of $f$, conveniently normalized. The formula
$(\ref{Weyl})$ can be obtained by a similar computation.
For that purpose, let us define the {\it symplectic Fourier
transformation} $\F_\Xi :\;\S'(\Xi)\rightarrow\S'(\Xi)$
by
\begin{equation*}
(\F_\Xi f)(\xi):=\int_\Xi d\eta\;e^{i\sigma(\xi,\eta)}f(\eta).
\end{equation*}
Now, for any function $f: \Xi \to \C$ belonging to the Schwartz space
$\S(\Xi)$, we set
\begin{equation}\label{incep}
\Op(f):=\int_\Xi d\xi\;(\F_\Xi^{-1} f)(\xi)\;\!W(\xi).
\end{equation}
By using (\ref{Waction}), one gets formula (\ref{Weyl}). Then it is
easy to verify that the relation $\Op(f) \Op(g)= \Op(f \circ g)$ holds
for $f,g\in\S(\Xi)$ if one uses the Moyal product introduced in
(\ref{Moyal}).

\subsection{Connection with $C^*$-dynamical systems and crossed
products}\label{schrorep}

One can turn to the crossed product formalism just by examining the
composition law $\circ$ in conjunction with a partial Fourier transformation.
It is more instructive to take the natural path and recover it by
reformulating the canonical commutation relations, so as to get a
dynamical system.

The family $\{V(p)\}_{p\in X^\star}$ is just part of
the functional calculus of the position operator $Q$,
defined by the Spectral Theorem. For any Borel function
$a:X\rightarrow\C$ one denotes by $a(Q)$ the normal (unbounded)
operator in $L^2(X)$ of multiplication by the function $a$. The
formula (\ref{WCCR}) is just a particular case of the more general one
\begin{equation*}
U(-x)\;\!a(Q)\;\!U(x)=a(Q+x),\qquad x\in X.
\end{equation*}
Usually one works with $a$'s belonging to some $C^*$-algebra $\A$ of
continuous functions on $X$ and $a\to a(Q)$ becomes a representation of $\A$.
In its turn, $U$ is a unitary, strongly continuous group representation.
Let us now point out a general framework encompassing this situation.

\begin{definition}\label{DS}
{\rm A $C^*$-{\it dynamical system} is a triple $(\A,\theta,X)$ formed
by a locally compact group $X$, a $C^*$-algebra $\A$ and a group
morphism $\theta:X\rightarrow\Aut(\A)$ of $X$ into the group of
automorphisms of $\A$ which is continuous in the sense that for any
$a\in\A$, the map $X\ni x\mapsto\theta_x(a)\in\A$ is continuous.}
\end{definition}

\begin{definition}
{\rm A {\it covariant representation} of the $C^*$-dynamical system
$(\A,\theta,X)$ is a triple $(\H,r,T)$, where $\H$ is a (separable)
Hilbert space, $r:\A \to \B(\H)$ a non-degenerate $({}^*)$-representation
of $\A$ in $\H$ and $T:X\rightarrow\U(\H)$ a strongly continuous unitary
representation of $X$ in $\H$, such that for all $x\in X$ and
$a\in\A$ one has $T(x)r(a)T(-x)=r[\theta_x(a)]$.}
\end{definition}

It is quite obvious how to recover our Quantum Mechanical framework
from these general definitions. $X$ is the configuration space
$\R^N$ and $\theta$ is the action of $X$ by translations on some
suitable $C^*$-algebra $\A$ of functions on $X$: $[\theta_x(a)](y)
:=a(y+x)$. To be ``suitable'' $\A$ has to be stable under translations
($a\in\A\Rightarrow \theta_x(a)\in\A$, $\forall x\in X$) and composed
of bounded, uniformly continuous functions, because this ensures the
desired continuity of $\theta$. A covariant representation is
recovered by taking $\H=L^2(X)$, $T(x)=U(-x)=e^{+ix\cdot P}$
and $r(a)=a(Q)$. This natural representation is called {\it the
Schr\"odinger representation}. In a certain sense, the kinematics of
our quantum particle is described by two composed objects. First a
classical setting, consisting of a configuration space and a group
acting in a particular way on this configuration space, the action
being then raised to functions. Second, a quantum implementation of the
classical system in a Hilbert space by a suitable algebra of position
observables and a unitary representation of the group, the two
satisfying some natural commutation relations. We note that we are
quite close of the concept of {\it imprimitivity system}, often used
in the foundational theory of Quantum Mechanics; we refer to
\cite{Varadarajan}.

A digression: With any locally compact group $X$ (we shall assume it
abelian) one associates naturally {\it the group} $C^*$-{\it algebra}
$C^*(X)$. It is the completion in some suitable norm of the space
$L^1(X)$ (with respect to the Haar measure), which is a Banach
$^*$-algebra with the involution $\alpha^*(x):=\overline{\alpha(-x)}$
and the convolution product $(\alpha\star\beta)(x)=\int_Xdy\;\!\alpha(y)
\;\!\beta(x-y)$. The virtue of $C^*(X)$ consists in the fact that its
non-degenerate representations are in one-to-one correspondence with
the unitary representations of the group $X$. In one direction, if
$T:X\rightarrow\U(\H)$ is such a group representation, then
$\ST(\alpha):=\int_Xdx\;\!\alpha(x)\;\!T(x)$ is a representation of the
Banach $^*$-algebra $L^1(X)$, that extends to a representation of the
group $C^*$-algebra. Since the group $X$ was taken abelian, a Fourier
transformation realizes an isomorphism between $C^*(X)$ and the
$C^*$-algebra $\CO(X^\star)$ (with the pointwise operations and the sup-norm)
of all continuous functions vanishing at infinity on the dual
$X^\star$. By composing $\ST$ with the isomorphism one gets a
representation $\sigma_{\hbox{\tiny T}}$ of $\CO(X^\star)$.
In the particular case in which $X=\R^N$ is represented in
$L^2(X)$ by translations, $T(x)=e^{ix\cdot P}$,
this representation is exactly the functional calculus of the
momentum operator: $\sigma_{\hbox{\tiny T}}(b)=b(P)$, $\forall b
\in \CO(X^\star)$. We see now why the $L^1$-norm has to be replaced:
not only a $C^*$-norm has better technical properties, but, very
concretely, the norm of the operator $b(P)$ equals the sup-norm of
$b$ and not the $L^1$-norm of its Fourier transform.

After this digression, let us come back to the above $C^*$-dynamical
system $(\A,\theta,X)$, covariantly represented by some arbitrary
$(\H,r,T)$. We would like to define a single $C^*$-algebra containing
in some subtle sense both $\A$ and $C^*(X)\cong \CO(X^\star)$ and
taking also into account the action of $X$ by automorphisms of $\A$.
And, hopefully, the two maps composing the covariant representation
should be condensed into a single representation of this large
$C^*$-algebra. Very roughly, for our Quantum Mechanical problem, one
tries to put together position and momentum observables such that
their commutation rules are respected.
This was also our point of view on the pseudodifferential calculus
in Subsection {\bf \ref{sub11}}, but with a different presentation of
the commutation relations.
It comes out that the right construction is as follows:

(1) On the Banach space $L^1(X;\A)$ (with the natural norm) one
defines an involution $\phi^\d(x):=[\phi(-x)]^*$, $^*$~being the
involution in $\A$, and a composition law
\begin{equation}\label{azi}
(\phi\d\psi)(x):=\int_Xdy\;\theta_{(y-x)/2}\;\!\left[\phi(y)\right]\;\!
\theta_{y/2}\;\!\left[\psi(x-y)\right].
\end{equation}
Then $L^1(X;\A)$ becomes a Banach $^*$-algebra.

(2) One introduces the $C^*$-norm $\|\phi\|:=\sup\| \pi(\phi)\|$,
the ``sup'' being taken over all the non-degenerate
representations of $L^1(X;\A)$. The completion of $L^1(X;\A)$ with
respect to $\|\cdot\|$ is denoted by $\A\rtimes_\theta X$ and is
called {\it the crossed product of} $\A$ {\it by the action}
$\theta$ {\it of the group} $X$. Obviously this generalizes the
concept of group $C^*$-algebra, that we recover for $\A = \C$.

Let us point out that in order to recover the Weyl calculus and not
the Kohn-Nirenberg pseudodifferential calculus we have a slightly unusual
form for the product operation in the crossed product algebra. In fact this
is isomorphic with the usual one (see for example {\bf \ref{s3.1}}).

Now, for a given covariant representation $(\H,r,T)$, one sets
$r\rtimes T:\A\rtimes_\theta X\rightarrow \B(\H)$, uniquely
defined by the action on $L^1(X;\A)$:
\begin{equation*}
(r\rtimes T)(\phi):=\int_X dx\;r\left[\theta_{x/2}\big(\phi(x)\big)
\right]T(x).
\end{equation*}
This is, indeed, a non-degenerate representation of the
$C^*$-algebra $\A\rtimes_\theta X$. In fact there is also a converse
construction, hence there are no other non-degenerate representations.

In the special case ($X=\R^N$, action by translations on an
invariant $C^*$-algebra of bounded, uniformly continuous functions on
$X$) {\it the composition law $\d$ is isomorphic to the Moyal product}
(\ref{Moyal}). The isomorphism is just a partial Fourier
transformation, also transporting the involutions $^\d$ and $^\circ$
one into the other. If $(\H,r,T)$ is the Schr\"odinger representation,
then by composing $r\rtimes T$ with the isomorphism one gets the
mapping $\Op$. So, in this particular case, the crossed product is
just another form of the Weyl pseudodifferential calculus. Remark that
the possibility of choosing the algebra $\A$ will offers a
flexibility which was not evident  at the pseudodifferential level.
In the sequel we shall extend all these facts to a much larger
setting.

\section{Twisted crossed products}\label{s2}

This section is mainly dedicated to a brief summary of twisted
$C^*$-dynamical systems,
twisted crossed products and of their representations. We follow
the standard references \cite{Busby}, \cite{Raeburn1}, \cite{Raeburn2}
and \cite{Packer}. To simplify, we undertake various hypotheses which
are not needed for part of the arguments. Primarily, we assume that
an {\it abelian} locally compact group acts upon an {\it abelian}
$C^*$-algebra. This will favour later on the use of Fourier transforms
and of Gelfand theory. Instead of working in terms of universal
properties, we treat the twisted crossed product as the envelopping
$C^*$-algebra of a $L^1$-type Banach $^*$-algebra.
To make the transition towards pseudodifferential operators and
the magnetic case, we introduce at the
end of the present section a special type of twisted
crossed products, in which the algebra is composed of continuous
functions defined on the group. It is preceded and prepared by some
considerations in group cohomology. Our setting is that of Polish
modules, as in \cite{Moore}, but we use continuous and not Borel cochains,
so \cite{Guichardet} is also relevant for our framework.

\subsection{Twisted $C^*$-dynamical systems and their covariant
representations}\label{Ccovrep}

Let us start abruptly with the relevant definition and explain
afterwards its ingredients. Some of the explanations will be used
only later on.

\begin{definition}\label{totazi}
{\rm We call {\it abelian twisted} $C^*$-{\it dynamical system} a
quadruplet $(\A,\theta,\omega,X)$, where
\begin{itemize}
\item[(i)] $X$ is an abelian, second countable locally compact
group,
\item[(ii)] $\A$ is an abelian, separable $C^*$-algebra,
\item[(iii)] $\theta : X \rightarrow \Aut(\A)$ is a group morphism
from $X$ to the group of automorphisms of $\A$, such that
$x\mapsto\theta_x(a)$ is continuous for all $a\in\A$,
\item[(iv)] $\omega$ is a strictly continuous normalized
$2$-cocycle on $X$ with values in the unitary group of the multiplier
algebra of $\A$.
\end{itemize}}
\end{definition}

The couple $(\theta,\omega)$ is called {\it a twisted action of $X$
on $\A$}. Very often, we shall use the shorter expression {\it
twisted dynamical system} for the quadruplet $(\A,\theta,\omega,X)$.

{\bf Remarks:}

(A) Almost everything in this section would be true, with only some
slight modifications, without assuming $\A$ and $X$ to be
abelian. However, our main interest lies in the connection between
twisted dynamical systems and pseudodifferential
theories. And for this purpose commutativity is extremely useful,
almost essential. Therefore we do assume it from the very beginning.
Separability conditions are needed to remain as close as possible to standard
references (\cite{Busby}, \cite{Raeburn1}, \cite{Raeburn2}), but also to 
assign to $(\A,\theta,\omega,X)$ a ``perfect'' group cohomology in the sense 
of Moore (see \cite{Moore} and {\bf \ref{cohomol}}). We intend to weaken the
separability condition on $\A$ in a forthcoming publication.

(B) The $2$-cocycle, introduced at  (iv) and explained at (D),
has to be unitary-valued. In a non-unital $C^*$-algebra unitarity
makes no sense, which requires the introduction of the multiplier
algebra. We refer to \cite{Pedersen} or to \cite{Williams} for this
useful concept and recall only some simple facts. It is known that
any non-unital $C^*$-algebra $\A$ can be embedded into several
larger unital $C^*$-algebras as an essential ideal ($\A$ is
an essential ideal of $\M$ if $\A \cap \I \neq 0$ for all non-zero
closed ideals $\I$ in $\M$).
Among these algebras there is a largest one, unique up to
isomorphisms, called {\it the multiplier algebra} and denoted by
$\M(\A)$. It can be introduced nicely as the class of all double
centralizers of $\A$, but a concrete definition might be easier to
grasp. Since $\A$ can be represented faithfully in some Hilbert space
$\H$, we shall just imagine that $\A$ is a $C^*$-subalgebra of
$\B(\H)$.
Then one sets $\M(\A):=\{T\in\B(\H)\mid Ta,\;\!aT\in\A,\;
\forall a\in\A\}$. It can be shown that $\M(\A)$ is also
commutative and that a different faithful representation would lead to
some isomorphic copy of $\M(\A)$. Let us notice that if the
$C^*$-algebra $\A$ is already unital, then one has naturally
$\A=\M(\A)$. The $C^*$-algebra $\M(\A)$ is endowed with a second natural
topology, the topology generated by the family of seminorms
$m \mapsto \|ma\|$ for all $a \in \A$,
$\|\cdot\|$ being the norm of $\B(\H)$. This topology,
with respect to  which $\M(\A)$ is complete and contains $\A$ densely,
is called {\it the strict topology}.

(C) Since $\M(\A)$ is unital, we can consider the unitary group
$\U(\A)\equiv \U\M(\A):=\{m\in\M(\A)\mid m^*m=1\}$, simply called
{\it the unitary group} of $\A$. By restricting the strict topology
to $\U(A)$ we get a topological group. Since $\A$ was supposed
separable, its topology is Polish (metrizable, separable and complete).

(D) If one forgets about $\omega$, the remaining data $(\A,\theta,X)$
form a $C^*$-{\it dynamical system} (with plenty of extra assumptions)
according to \cite{Doplicher}, \cite{Pedersen} or \cite{Williams};
see also Definition \ref{DS}.
Let us now explain the point (iv) of Definition \ref{totazi}.
A $2$-cocycle is a function $\omega : X\times X\rightarrow \U(\A)$,
continuous with respect to the strict topology on $\U(\A)$, such that
for all $x,y,z\in X$~:
\begin{equation}\label{cocond}
\omega(x+y,z)\;\!\omega(x,y)=\theta_x[\omega(y,z)]\;\!\omega(x,y+z).
\end{equation}
We shall also assume it to be normalized:
\begin{equation}\label{condi2}
\omega(x,0)=\omega(0,x)=1,\qquad \hbox{for all }x \in X.
\end{equation}
It is known that any
automorphism of $\A$ extends uniquely to an automorphism of $\M(\A)$
and, obviously, leaves $\U(\A)$ invariant. By applying this fact to
$\theta_x$ and by denoting the extension with the same symbol, one
gives a sense to (\ref{cocond}). Actually, by suitable particularizations
in (\ref{cocond}), we get $\theta_{-x}[\omega(x,0)]=\omega(0,0)=\omega(0,x)$,
$\forall x\in X$, hence for normalization it suffices to ask $\omega(0,0)=1$.
The required continuity can be rephrased by saying that for any $a\in\A$,
the map $X\times X \ni(x,y)\mapsto a\omega(x,y)\in\A$ is continuous.
In fact Borel conditions could be imposed instead of continuity for most 
of the constructions and results; we do not pursue this here.
``$2$-cocycle'' is a concept belonging to group cohomology. We shall give 
further details in Subsection {\bf \ref{cohomol}}.

Let us now discuss some special features due to the fact that we
assume $\A$ abelian.

(E) Gelfand theory describes completely the structure of {\it abelian}
$C^*$-algebras. We first note that if $S$ is a locally compact
space, then $\CO(S):=\{a:S\rightarrow\C\mid a \text{ continuous, }
a\rightarrow 0 \text{ when }x\rightarrow\infty\}$ is an abelian
$C^*$-algebra with the operations defined pointwise and the
$\sup$-norm. $\CO(S)$ has a unit if and only if $S$ is compact.
Actually {\it all abelian $C^*$-algebras are of this form}. We define
{\it the Gelfand spectrum} $\SA$ of $\A$ to be the family of
all characters of $\A$ ({\it a character} is just a morphism
$\nu:\A\rightarrow \C$). With the topology of simple convergence
$\SA$ is a locally compact space, which is compact exactly when
$\A$ is unital. And the mapping $\G:\A\rightarrow \CO(\SA)$ given
by $[\G(a)](\nu):=\nu(a)$, for $a\in\A$ and $\nu \in \SA$ is
an isomorphism.

(F) If the $C^*$-algebra $\A\cong \CO(\SA)$ is not unital, then
$\BC(\SA)$, the $C^*$-algebra of all bounded and continuous
complex functions on $\SA$, surely is. It contains
$\CO(\SA)$ as an essential ideal.
In fact $\BC(\SA)$ can be identified with the multiplier algebra
$\M(\A)$ of $\A$ (see \cite{Williams}). Thus the unitary group of
$\A$ is identified with $\CC(\SA,\T)$, the family of all
continuous functions on $\SA$ taking values in the group
$\T$ of complex numbers of absolute value~$1$. Moreover, the strict
topology on $\CC(\SA,\T)$ coincides with the topology of uniform
convergence on compact subsets of $\SA$.

(G) Abelian $C^*$-algebras being so special $\big($cf.~(E)$\big)$, the
corresponding $C^*$-dynamical systems are also special. They are in
fact given by topological dynamical systems and this explains the
terminology. The central remark is that the only automorphisms of
$\CO(\SA)$ are those implemented by homeomorphisms of the
underlying locally compact space $\SA$. Here ``implementation''
means just composition of the elements of $\CO(\SA)$ with the
homeomorphism. Thus $\theta$ induces an action of $X$ through
homeomorphisms of $\SA$.

The coherent way to represent a twisted dynamical system in a Hilbert
space is given by

\begin{definition}\label{RCT}
{\rm Given a twisted dynamical system $(\A,\theta,\omega,X)$, we call {\it
covariant representation} a Hilbert space $\H$ together with two maps
$r:\A\rightarrow \B(\H)$ and $T:X\rightarrow \U(\H)$ satisfying
\begin{itemize}
\item[(i)] $r$ is a non-degenerate representation,
\item[(ii)] $T$ is strongly continuous and $\
T(x)T(y)=r[\omega(x,y)]T(x+y), \quad \forall x,y\in X$,
\item[(iii)] $T(x)r(a)T(x)^*=r[\theta_x(a)], \quad \forall x\in X,\;
a\in\A$.
\end{itemize}}
\end{definition}

One observes that $T$ is a sort of generalized projective
representation of $X$. The usual notion of projective representation
corresponds to the case in which for all $x,y\in X$, $\omega(x,y)\in
\T$, i.e.~``$\omega(x,y)$ is a constant function on the spectrum
$\SA$ of $\A$''. Let us already mention that constant magnetic fields
lead to such a situation.
Condition (iii) says that at a represented level the
automorphism $\theta_x$ is implemented by the unitary equivalence
associated to $T(x)$. It could also be interpreted as an a priori
prescribed commutation rule between the elements of the group and the
elements of the algebra when they are put together in $\B(\H)$ by
representations.

\subsection{Twisted crossed products and their representations}

Let $(\A,\theta,\omega,X)$ be a twisted dynamical system.
We start by mixing together the algebra $\A$ and the space $L^1(X)$
in a way to form a Banach $^*$-algebra. We define $L^1(X;\A)$,
the Bochner integrable equivalence classes of $\A$-valued functions 
(with respect to the Haar measure), endowed with the norm
$\n\phi\n:=\int_Xdx\;\|\phi(x)\|_{\A}$. Let us also fix an element $\tau$
of the set $\End(X)$ of continuous endomorphisms of $X$. Particular cases
are $\0,\1 \in\End(X)$, $\0(x):=0$ and $\1(x):=x$, for all $x\in X$.
Addition and substraction of endomorphisms are well-defined. For elements
$\phi,\psi$ of $L^1(X;\A)$ and for any point $x\in X$ we set
\begin{equation}\label{produit}
(\phi\d^\omega_\tau \psi)(x):=\int_Xdy\;\!\theta_{\tau(y-x)}\left[\phi(y)
\right]\;\!\theta_{(\1-\tau)y}\left[\psi(x-y)\right]\;\!\theta_{-\tau x}
\left[\omega(y,x-y)\right]
\end{equation}
and ($a^*$ is the adjoint of $a$ in $\A$)
\begin{equation}\label{invo}
\phi^{\d_{\tau}^\omega}(x):=\theta_{-\tau x}[\omega(x,-x)^{-1}]
\;\!\theta_{(\1-2\tau)x}\left[\phi(-x)^*\right].
\end{equation}
The expression (\ref{invo}) becomes much simpler if $\omega(x,-x)=1$,
which will be the case in most of the applications.

\begin{lemma}
For two functions $\phi$ and $\psi$ in $L^1(X;\A)$ and for $\tau \in
\End(X)$, the function $\phi\d^\omega_\tau \psi$ belongs to
$L^1(X;\A)$. With the composition law $\d^\omega_\tau$ and the
involution $^{\diamond_\tau^\omega}$, $L^1(X;\A)$ is a Banach
$^*$-algebra. These Banach $^*$-algebras are isomorphic for different
$\tau$'s.
\end{lemma}

\begin{proof} The fact that $L^1(X;\A)$ is stable under
$\d^\omega_\tau$ follows from the relations
\begin{equation*}
\begin{array}{l}
\|\theta_{\tau(y-x)}\left[\phi(y)\right]\theta_{(\mathfrak 1-\tau)y}
\left[\psi(x-y)\right]\theta_{-\tau x}\left[\omega(y,x-y)\right]
\|_{\A}\leq \|\phi(y)\|_{\A}\|\psi(x-y)\|_{\A},\\\\
\int_Xdx\;\|(\phi\d_\tau^\omega \psi)(x)\|_{\A}\leq
\int_Xdx\int_Xdy\;\|\phi(y)\|_{\A}\|\psi(x-y)\|_{\A}=\n\phi\n \n\psi\n.
\end{array}
\end{equation*}
\noindent
The associativity of this composition law is easily deduced from
the 2-cocycle property of $\omega$. All the other requirements also
follow by routine calculations.

The isomorphisms are the mappings
$$ m_{\tau,\tau'}:L^1(X;\A)\rightarrow L^1(X;\A), \quad
\left(m_{\tau,\tau'}\phi\right)(x):=\theta_{(\tau'-\tau)x}[\phi(x)],
\qquad x\in X.$$
On the first copy of $L^1$ one considers the structure defined by $\tau'$
and on the second that defined by $\tau$. Note the obvious relations
$m_{\tau,\tau'}m_{\tau',\tau''}= m_{\tau,\tau''}\ $ and
$\ [m_{\tau,\tau'}]^{-1}=m_{\tau',\tau}$
for all $\ \tau,\tau',\tau''\in \End(X)$.
\end{proof}

One finds in the literature only the case $\tau=\0$. We introduced all
these isomorphic structures because they help in understanding
$\tau$-quantizations in pseudodifferential theory.

We recall that {\it a $C^*$-norm} on a $^*$-algebra has to satisfy
$\|a^*a\|=\| a\|^2$. Since $C^*$-norms have many technical advantages
and since $\n\cdot\n$ has not this $C^*$-property, we shall make now
some adjustments, valid in an abstract setting. A Banach $^*$-algebra
$\CC$ with norm $\n \cdot \n$ is called {\it an $A^*$-algebra} when
it admits a $C^*$-norm or, equivalently, when it has an injective 
representation in a Hilbert space \cite{Tak}. 
In this case we can consider {\it the standard $C^*$-norm} on
it, defined as the supremum of all the $C^*$-norms, that we shall
denote by $\|\cdot\|$. A rather explicit formula for $\|\cdot\|$ is
$\|b\|=\sup\{\|\pi(b)\|_{\B(\H)}\;|\;(\pi,\H)\text{ is a
representation}\}$. One has $\|\phi\|\leq\n\phi\n$ for all $\phi$.
The completion with respect to this norm will be a $C^*$-algebra
containing $\CC$ as a dense $^*$-subalgebra. We call it {\it the
envelopping $C^*$-algebra of $\CC$}. It is known that
$\left(L^1(X;\A),\d^\omega_\tau,^{\d_\tau^\omega},\n\cdot\n\right)$
is indeed an $A^*$-algebra.

\begin{definition}
{\rm The envelopping $C^*$-algebra of $\left(L^1(X;\A),\d^\omega_\tau,
{}^{\d_\tau^\omega},\n \cdot \n\right)$ will be called {\it the twisted
crossed product of $\A$ by $X$ associated with the twisted action
$(\theta,\omega)$ and the endomorphism} $\tau$. It will be denoted by
$\A\rtimes^\omega_{\theta,\tau} X$.}
\end{definition}

The $C^*$-algebra $\A\rtimes^\omega_{\theta,\tau} X$ has a rather
abstract nature. But most of the time one uses efficiently the
fact that $L^1(X;\A)$ is a dense $^*$-subalgebra, on which everything
is very explicitly defined. Let us even observe that the algebraic
tensor product $\A\odot L^1(X)$ may be identified with the dense
$^*$-subspace of $L^1(X;\A)$ (hence of
$\A\rtimes^\omega_{\theta,\tau} X$ also) formed of functions with
finite-dimensional range. The isomorphism $m_{\tau,\tau'}$
extends nicely to an isomorphism from
$\A\rtimes^\omega_{\theta,\tau'}X$ to $\A\rtimes^\omega_{\theta,\tau}X$.

The next lemma shows clearly the importance of twisted crossed
products as a way to bring together the informations contained in a
twisted dynamical system.

\begin{lemma}
If $(\H,r,T)$ is a covariant representation of $(\A,\theta,\omega,X)$
and $\tau \in \End(X)$, then $r\rtimes_\tau T$ defined on $L^1(X;\A)$
by
\begin{equation*}
(r\rtimes_\tau T)\phi:=\int_Xdx\,r\left[\theta_{\tau x}
\big(\phi(x)\big)\right]T(x)
\end{equation*}
extends to a representation of $\A\rtimes^\omega_{\theta,\tau} X$,
called {\rm the integrated form of} $(r,T)$. One has
$r\rtimes_{\tau'} T=\left(r\rtimes_{\tau}T\right)\circ
m_{\tau,\tau'}$ if $\tau,\tau'\in\End(X)$.
\end{lemma}

\begin{proof}
Some easy computations show that $r\rtimes_\tau T$ is a
representation of the Banach $^*$-algebra $\left(L^1(X;\A),
\d^\omega_\tau,^{\d_\tau^\omega}\right)$. Then, by taking
into account the formula for $\|\cdot\|$, one gets
$\|(r\rtimes_\tau T)\phi\|_{\B(\H)}\;\leq\;\|\phi\|$, $\;
\forall \phi\in L^1(X;\A)$. Thus $r\rtimes_\tau T$ extends to
$\A\rtimes^\omega_{\theta,\tau} X$ by density and, by approximation,
this extension has all the required algebraic properties.

The relation $r\rtimes_{\tau'} T=\left(r\rtimes_{\tau}T\right)\circ
m_{\tau,\tau'}$ is checked readily on $L^1(X;\A)$ and obviously
extends to the full twisted crossed product.
\end{proof}

We note that one can recover the covariant representation from
$r\rtimes_\tau T$. Actually, there is a one-to-one correspondence
between covariant representations of a twisted dynamical system and
non-degenerate representations of the twisted crossed product.
This correspondence preserves equivalence, irreducibility and direct
sums. We do not give explicit formulae, since we do not use them.

Finally, we would like to show that $\A\rtimes^\omega_{\theta,\tau} X$
is generated in some sense (by using representations) by $\A$ and the
$L^1$-space of the group $X$. The next result was included without
proof in \cite{Purice1}. For completeness, we prove it here by using
the approach of \cite{Georgescu1} for the untwisted case.

\begin{proposition}\label{concreta}
Let $(\H,r,T)$ be a covariant representation of the twisted
dynamical system $(\A,\theta,\omega,X)$.
\begin{itemize}
\item[{\rm (a)}] The mapping $\ST:L^1(X)\rightarrow \B(\H)$ given by
$\ST(b):=\int_X dx\;b(x)T(x)$ is a linear contraction,
\item[{\rm (b)}] The norm closure of the vector space generated by
$\{r(a)\ST(b)\;|\; a\in\A,\;\!b\in L^1(X)\}$ is equal to the norm
closure of the vector space generated by $\{\ST(b)r(a)\;|\;
a\in\A,\;\!b\in L^1(X)\}$. Furthermore, both coincide with the
$C^*$-algebra $(r\rtimes_{\tau} T)
\left(\A\rtimes^\omega_{\theta,\tau} X\right)$ for any
$\tau \in \End(X)$.
\end{itemize}
\end{proposition}

\begin{proof}
The proof of statement (a) is obvious. The first part of statement
(b) is a simple corollary of the following claim: For any $b\in
L^1(X)$, any $a \in \A$ and any $\e >0$ there exist $\{x_k\}_{k=1}^L$
with $x_k \in X$ and $\{b_k\}_{k=1}^L$ with $b_k \in L^1(X)$ such
that
\begin{equation}\label{diff}
\left\| \ST(b)r(a)-\sum_{k=1}^L r[\theta_{x_k}(a)]\ST(b_k)
\right\|\le \e.
\end{equation}
To prove this, consider a finite set $\{J_k\}_{k=0}^L$ of continuous
functions on $X$ such that $0 \leq J_k \leq 1$ and
$\sum_{k=0}^L J_k=1$. The functions $J_k$ will have a compact support
for $k \geq 1$. Let also $\{x_k\}_{k=1}^L$ be a finite
set of elements of $X$. Then one has
\begin{eqnarray*}
\ST(b)r(a) & = & \sum_{k=1}^L r[\theta_{x_k}(a)] \int_X d x
J_k(x)b(x)T(x) + \sum_{k=1}^L \int_X d x\big(r[\theta_{x}(a)]-
r[\theta_{x_k}(a)]\big) J_k(x)b(x)T(x) \\
&& + \int_X d x\; r[\theta_x(a)] J_0(x)b(x)T(x).
\end{eqnarray*}
By setting $b_k := b J_k \in L^1(X)$, the l.h.s.~term of
(\ref{diff}) is less or equal to
\begin{equation}\label{maj}
\max_{k=1,\dots, L}\sup_{x \in \hbox{\small supp}\;\! J_k}\|
\theta_x(a) - \theta_{x_k}(a)\|\;\|b\|_{L^1} + \|a\|
\int_{\hbox{\small supp}\;\! J_0} d x |b(x)|,
\end{equation}
Let $K$ be a compact subset of $X$ such that
$\int_{X\setminus K}d x |b(x)| \leq \frac{\e}{2\|a\|}$, and let $V$
be a neighbourhood of $0$ in $X$ such that
$\sup_{x \in V} \|\theta_x(a)-a\| \le \frac{\e}{2 \|b\|_{L^1}}$.
We now choose the set $\{x_k\}_{k=1}^L$ such that $K \subset
\cup_{k=1}^L (x_k + V)$, and the collection $\{J_k\}_{k=0}^L$ such
that $\hbox{supp}\;\! J_k \subset x_k +V$ for
$k \in \{1,\ldots,L\}$ and $\hbox{supp}\;\! J_0 \subset X \setminus K$.
The inequality (\ref{diff}) is then easily obtained from (\ref{maj}).

For the final statement of (b), let us observe that $(r \rtimes_\tau T)
\left[m_{\tau,\mathfrak 0}(a \otimes b)\right]= r(a)\ST(b)$ for any
$a \in \A$ and any $b\in L^1(X)$. The conclusion follows from the density
of $\A\odot L^1(X)$ in $L^1(X;\A)$.
\end{proof}

\subsection{Group cohomology}\label{cohomol}

We recall some definitions in group cohomology. They will be
used in the next sections to show that standard matters as gauge
invariance and $\tau$-quantizations have a cohomological flavour.
Now they will serve to isolate twisted dynamical systems for which
a generalization of the Schr\"odinger representation exists.

Let $X$ be an abelian, locally compact group and $\U$ a Polish
abelian group. Recall that a Polish group is a group with a
compatible metrizable, separable and complete topology.
In our applications $\U$ will usually not be locally compact, being
the unitary group of an abelian $C^*$-algebra, as in Subsection
{\bf \ref{Ccovrep}}. We also assume that $\U$ is an $X${\it -module},
i.e.~that there exists a continuous action $\theta$ of $X$ by
automorphisms of $\U$. We shall use for $X$ and $\U$ additive and
multiplicative notations, respectively.

The class of all continuous functions $:X^n\rightarrow\U$ is denoted
by $C^n(X;\U)$; it is obviously an abelian group (we use once again
multiplicative notations). Elements of $C^n(X;\U)$ are called
{\it (continuous) $n$-cochains}.  For any $n\in\N$, we define
{\it the coboundary map}
$\delta^n:C^n(X;\U)\rightarrow C^{n+1}(X;\U)$ by
\begin{equation*}
\left[\delta^n(\rho)\right](x_1,\dots,x_n,x_{n+1}):=
\theta_{x_1}\left[\rho(x_2,\dots,x_{n+1})\right]
\prod_{j=1}^n\rho(x_1,\dots,x_j+x_{j+1},\dots,x_{n+1})^{(-1)^j}
\rho(x_1,\dots,x_{n})^{(-1)^{n+1}}.
\end{equation*}
It is easily shown that $\delta^n$ is a group morphism and that
$\delta^{n+1}\circ\delta^n=1$ for any $n\in\N$. It follows
that $\text{im}(\delta^n)\subset\text{ker}(\delta^{n+1})$.

\begin{definition}{\rm
\begin{itemize}
\item[(a)] $Z^n(X;\U):=\text{ker}(\delta^{n})$ is
called {\it the set of $n$-cocycles (on $X$, with coefficients
in $\U$)}.
\item[(b)] $B^n(X;\U):=\text{im}(\delta^{n-1})$ is called {\it the set
of $n$-coboundaries}.
\end{itemize}
}
\end{definition}

One notices that $Z^n(X;\U)$ and $B^n(X;\U)$ are subgroups of
$C^n(X;\U)$, and that $B^n(X;\U)\subset Z^n(X;\U)$.

\begin{definition}
{\rm The quotient $H^n(X;\U):=Z^n(X;\U)/B^n(X;\U)$ is called
{\it the $n$'th group of cohomology (of $X$ with coefficients in
$\U$)}. Its elements are called {\it classes of cohomology}.}
\end{definition}

We shall need only the cases $n=0,1,2$, which we outline now for the
convenience of the reader.
For $n=0$, parts of the definitions are simple conventions. We
set $C^0(X;\U):=\U$. One has
$\left[\delta^0(a)\right](x)=\frac{\theta_x(a)}{a}$, $\forall a\in\U,
x\in X$. This implies that $Z^0(X;\U) =
\{a\in\U \; |\; a\text{ is a fixed point}\}$. By convention,
$B^0(X;\U)=\{1\}$.

The mapping $\delta^1:C^1(X;\U)\rightarrow C^2(X;\U)$ is given by
$\left[\delta^1(\lambda)\right](x,y)=\frac{\lambda(x)
\theta_x[\lambda(y)]}{\lambda(x+y)}$.
Thus a $1$-cochain $\lambda$ is in $Z^1(X;\U)$ if it is {\it a crossed
morphism}, i.e.~if it satisfies
$\lambda(x)\theta_x[\lambda(y)]=\lambda(x+y)$ for any $x,y\in
X$. Particular cases are the elements of $B^1(X;\U)$ (called {\it
principal morphisms}), those of the form
$\lambda(x)=\frac{\theta_x(a)}{a}$ for some $a\in\U$.

For $n=2$ one encounters a situation which was already taken into
account in the definition of twisted dynamical systems. The formula
for the coboundary map is
$$\left[\delta^2(\omega)\right](x,y,z)=\theta_x[\omega(y,z)]
\omega(x+y,z)^{-1}\omega(x,y+z)\omega(x,y)^{-1}.$$
Thus a $2$-cocycle is just a function satisfying the relation
(\ref{cocond}). $B^2(X;\U)$ is composed of $2$-cocycles of the form
$\omega(x,y)=\frac{\lambda(x)\theta_x[\lambda(y)]}{\lambda(x+y)}$ for
some $1$-cochain $\lambda$.

We are mainly interested in the case of $X$-modules coming from
$C^*$-dynamical systems, as in Subsection {\bf \ref{Ccovrep}},
the group $\U$ being the unitary group of some abelian $C^*$-algebra.
Our developments will need especially the case of algebras
$\A$ of continuous functions defined on the group itself. The next
result will be extremely significant for our formalism. For $n=2$, it
is a continuous version of Lemma 5.1 of \cite{Georgescu2}.
Recall that $\U:=\CC(X;\T)$, endowed with the strict topology, can be
interpreted as the unitary group associated with the $C^*$-algebra
$\CO(X)$.

\begin{lemma}\label{triv}
For $n \geq 1$, $H^n\big(X;\CC(X;\T)\big)=\{1\}$.
\end{lemma}

\begin{proof}
Let $\rho^n\in Z^n\big(X;\CC(X;\T)\big)$, i.e.~$\rho^n$ is a continuous
$n$-cochain satisfying for any $y_1,\dots,y_{n+1}\in X$
\begin{equation*}
\theta_{y_1}\left[\rho^n(y_2,\dots,y_{n+1})\right]
\prod_{j=1}^n\rho^n(y_1,\dots,y_j+y_{j+1},\dots,y_{n+1})^{(-1)^j}
\rho^n(y_1,\dots,y_{n})^{(-1)^{n+1}}=1.
\end{equation*}
We set in this relation $y_1=q$, $y_j=x_{j-1}$ for $j\ge 2$ and rephrase
it as
\begin{eqnarray}
\nonumber & \theta_{q}\left[\rho^n(x_1,\dots,x_{n})\right] = & \\
\nonumber & \rho^n(q+x_1,x_2,\dots,x_n)\prod_{j=1}^{n-1}
\rho^n(q,x_1,\dots,x_j+x_{j+1},\dots,
x_n)^{(-1)^{j}}\rho^n(q,x_1,\dots,x_{n-1})^{(-1)^n}, &
\end{eqnarray}
which is an identity in $\CC(X;\T)$.
One calculates both sides at the point $x=0$ and obtain
\begin{eqnarray}
\nonumber & \left[\rho^n(x_1,\dots,x_{n})\right](q) = & \\
\nonumber & \left[\rho^n(q+x_1,x_2,\dots,x_n)\right](0)\prod_{j=1}^{n-1}
\left[\rho^n(q,x_1,\dots,x_j+x_{j+1},\dots,
x_n)^{(-1)^j}\right](0)\left[\rho^n(q,x_1,\dots,x_{n-1})^{(-1)^n}\right](0). &
\end{eqnarray}
This means exactly $\rho^n=\delta^{n-1}(\rho^{n-1})$ for
\begin{equation}\label{psetri}
\left[\rho^{n-1}(z_1,\dots,z_{n-1})\right](q):=\left[
\rho^n(q,z_1,\dots,z_{n-1})\right](0)
\end{equation}
and thus any n-cocyle is at least formally an n-coboundary.

We show now that $\rho^{n-1}$ has the
right continuity properties. Let us recall that if $\CC(X;\T)$ is
endowed with the topology of uniform convergence on compact sets of
$X$ and if $Y$ is a locally compact space, then $\CC\big(Y;\CC(X;\T)\big)$
can naturally
be identified with $\CC(X\times Y;\T)$ (the proof of this statement is
an easy exercice). So $\rho^n$ can be interpreted as an element of
$\CC(X \times X^n;\T)$. Being obtained from $\rho^n$ by a
restriction  $\rho^{n-1}$ belongs to $\CC(X^n ;\T)$, and thus can be
interpreted as an element of $\CC\big(X^{n-1};\CC(X;\T)\big)\equiv
C^{n-1}\big(X;\CC(X;\T)\big)$, which finishes the proof.
\end{proof}

\begin{definition}
{\rm Let $\U$ be an abelian Polish $X$-module with action $\theta$
and $\omega\in Z^2(X;\U)$. We say that $\omega$ is {\it
pseudo-trivial} if there exists
another abelian Polish $X$-module $\U'$ with action $\theta'$
such that $\U$ is a subgroup of $\U'$, for each $x\in X$ one has
$\theta_x=\theta'_x\vert_\U$ and $\omega\in B^2(X;\U')$.}
\end{definition}

Thus, to produce pseudo-trivial $2$-cocycles, one has to find some
$\omega\in B^2(X;\U')$ such that $\omega(x,y)\in\U \subset \U'$ for any
$x,y\in X$ and such that $(x,y)\mapsto\omega(x,y)\in\U$ is continuous
with respect to the topology of $\U$. This is possible in principle
because the product
$\lambda(x)\theta_x[\lambda(y)][\lambda(x+y)]^{-1}$ can be
better-behaved than any of its factors. The particular choice
$[\lambda(z)](q)=[\omega(q,z)](0)$ we made in the proof of Lemma \ref{triv}
will lead in {\bf \ref{s4.1}} to the physicists' familiar transversal gauge.

Let us emphasize that most of the time pseudo-triviality cannot
be improved to a bona fide triviality. Very often, all the functions
$\lambda$ for which one has $\omega=\delta^1(\lambda)$ do not take all
their values in $\U$ or miss the right continuity. We shall outline
such a situation in the next subsection.

\subsection{Standard twisted crossed products}\label{sstand}

When trying to transform the formalism of twisted crossed products into
a pseudodifferential theory, one has to face the possible absence of
an analogue of the Schr\"odinger representation and this would lead us
too far from the initial motivation. The existence of a generalized
Schr\"odinger representation is assured by the pseudo-triviality of
the $2$-cocycle, and thus we restrict ourselves to a specific class
of twisted dynamical systems. In the same time we also restrict to
algebras $\A$ of complex continuous functions on $X$. This also is not
quite compulsory for a pseudodifferential theory, but it leads to a
simple implementation of pseudo-triviality (by Lemma \ref{triv}) and
covers easily the important magnetic case.

\begin{definition}
{\rm Let $X$ be an abelian, locally compact group. We call $X$-{\it
algebra} a $C^*$-algebra of bounded, uniformly continuous functions on
$X$, stable by translations: $\theta_x(a):=a(\cdot+x)\in\A$ for all
$a\in\A$ and $x\in X$.}
\end{definition}

The $C^*$-algebra $\BC_u(X):=\{a:X\rightarrow\C\mid a\text{\ is bounded
and uniformly continuous}\}$ is the largest one on which the action
$\theta$ of translations with elements of $X$ is norm-continuous.
But we shall denote by $\theta_x$ even the $x$-translation on $\CC(X)$,
the ${}^*$-algebra of all continuous complex functions on $X$ (which
is not a normed algebra if $X$ is not compact). The restriction of
$\theta_x$ on $\BC(X)$ is only strictly continuous.

Let us fix an $X$-algebra $\A$ with spectrum $\SA$. It will be technically
useful to note the existence of a continuous map
$\,\delta^\A:X\rightarrow \SA$ with a dense range . One sets 
$\delta_x^\A:\A\rightarrow \mathbb C$,
$\delta^\A_x(a):=a(x)$ and the continuity is obvious. The fact that it has
a dense range follows from the identification of the multiplier algebra
$\mathfrak M(\A)$ with a $C^*$-subalgebra of $\B\CC(X)$ (cf.~the proof of
Proposition \ref{Couchepin}) and a simple argument with Stone-\v Cech
compactifications. Remark that $\delta^\A$ is injective exactly when
$\CC_0(X)\subset\A$. If in addition $\A$ is unital, $\SA$ is a
compactification of $X$. Now the Gelfand isomorphism
$\A\cong \CC_0(\SA)$ can be put in very concrete terms: a bounded continuous
function $a:X\rightarrow \mathbb C$ belongs to $\A$ if and only if there
exists a (necessarily unique) function $\tilde a\in\CC_0(\SA)$ such that
$\tilde a\circ\delta^\A=a$. One has a similar criterion for $\M(\A)$ with
$\tilde{a} \in \B\CC(\SA)$.

\begin{definition}\label{tipA}
{\rm A function $b:X\rightarrow\mathbb C$ is {\it of type} $\A$ if
there exists $\tilde b\in\CC(\SA)$ such that $\tilde b\circ\delta^\A=b$.}
\end{definition}

If $\A$ is an $X$-algebra, then $(\A,\theta,X)$ is a $C^*$-dynamical
system. If we twist it, we get

\begin{definition}
{\rm {\it A standard twisted dynamical system} is a twisted dynamical
system $(\A,\theta,\omega,X)$ for which $\A$ is an $X$-algebra. The
$C^*$-algebra $\A\rtimes^\omega_{\theta,\tau}X$ is called {\it a standard
twisted crossed product}.}
\end{definition}

\begin{proposition}\label{Couchepin}
If $(\A,\theta,\omega,X)$ is a standard twisted dynamical system, then
$\omega$ is pseudo-trivial.
\end{proposition}

\begin{proof}
We shall prove the triviality of the 2-cocyle in the
unitary group of the $C^*$-algebra $\CO(X)$, i.e.~in $\CC(X;\T)$ endowed
with the strict topology.
We first show that the multiplier algebra of $\A$ can be identified
with a $C^*$-subalgebra of $\BC(X)$, the multiplier algebra of
$\CO(X)$. One remarks that, the trivial case $\A = \{0\}$ excluded,
the invariance of $\A$ under translations implies the non-degeneracy
of the natural faithful representation of $\A$ in $L^2(X)$. We keep
the same notation for $\A$ and for its image in $\B\big(L^2(X)\big)$.
It follows that $\M(\A)$ is contained in the double commutant of $\A$
(cf.~\cite{Fell}), which itself is contained in $L^{\infty}(X)$
(also represented in $L^2(X)$ by multiplication operators).
Moreover, any element $m$ of $\M(\A)$ is also a continuous map on
$X$: If $m$ would not be continuous in $x_0$, one could find by
translational invariance some element $a$ of $\A$ which is not
vanishing in a neighbourhood of $x_0$ and this obviously makes $ma
\in \A$ impossible. Thus $\M(\A) \subset \BC(X)$.

Let us now observe that $\U(\A)$ is a bounded subset of $\BC(X)$, and on 
bounded subsets of $\BC(X)$ the strict topology coincides with the
topology of uniform convergence on compact subsets of $X$. Similarly the
strict topology on bounded subsets of $\M(\A)$ coincides with the topology of
uniform convergence on compact subsets of $\SA$. But $\delta^{\A}: X \to \SA$
being continuous we deduce that on $\U(\A)$ the strict topology of $\M(\A)$
induces a finer topology than $\BC(X)$.

Thus $\U(\A)$ is naturally identified with a subgroup of $\CC(X;\T)$,
and the strict topology on $\U(\A)$ is finer than the strict topology
of $\CC(X;\T)$. $\omega$ can hence be considered as
an element of $Z^2\big(X;\CC(X;\T)\big)$, which coincides with
$B^2\big(X;\CC(X;\T)\big)$ by Lemma \ref{triv}, and this finishes the proof.
\end{proof}

{\bf Example.} The simplest $X$-algebra is composed only of constant functions.
If one takes $\A=\C$ then $\theta_x$ is the identity for all $x$
and $\omega:X\times X\rightarrow\T$ is sometimes called
{\it a multiplier of the group} $X$. Then
$\C\rtimes^{\omega}_{\text{id},\tau}X$ does no longer depend on $\tau$;
it is denoted by $C^*_\omega(X)$ and called {\it the twisted $C^*$-algebra
of the group $X$ associated with} $\omega$. Its non-degenerate representations
are in one-to-one correspondence with the $\omega$-projective
representations of $X$. We know from Proposition \ref{Couchepin} that
$\omega$ will be trivial if we enlarge $\mathbb T$ to $\CC(X;\T)$.
But $\omega\in B^2(X;\mathbb T)$ if and only if it is symmetric
($\omega(x,y)=\omega(y,x)$, $\forall x,y\in X$); this is proved in
\cite{Kleppner1}. Since for many groups $X$ other non-symmetric multipliers
are available, we see that very often triviality cannot be achieved within
$\T$. We shall encounter examples later on; in our
setting they correspond roughly to constant magnetic fields for $X=\R^n$
and give ``non-commutative tori'' for $X=\Z^N$.

\begin{remark}\label{remarca}
{\rm If $\omega, \omega'$ are two cohomologous elements of 
$Z^2\big(X;\U(\A)\big)$, i.e.~$\omega=\delta^1(\lambda)\:\!\omega'$ 
for some $\lambda\in C^1\big(X;\U(\A)\big)$, then the $C^*$-algebras
$\A\rtimes_{\theta,\tau}^{\omega}X$ and $\A\rtimes_{\theta,\tau}^{\omega'}X$ 
are naturally isomorphic: on $L^1(X;\A)$ the isomorphism is given by 
$\left[i^\lambda_\tau(\phi)\right](x) :=\theta_{-\tau x}[\lambda(x)]\phi(x)$. 
Thus $\CC_0(X)\rtimes_{\theta,\tau}^{\omega}X$ does not depend on $\omega$; this
will be strengthened in Proposition \ref{dada} (b). However this does not work
if $\lambda$ only belongs to $C^1\big(X;\CC(X;\T)\big)$ and $\A$ is not
$\CC_0(X)$; in general $\theta_{-\tau x}[\lambda(x)]\phi(x)$ gets out of $\A$
and $i^\lambda_\tau$ is no longer well-defined. For $\omega$ and $\omega'$
defining different classes of cohomology, $\A\rtimes_{\theta,\tau}^{\omega}X$
and $\A\rtimes_{\theta,\tau}^{\omega'}X$ are in general different $C^*$-algebras.}
\end{remark}

In the sequel we fix a standard twisted dynamical system
$(\A,\theta,\omega,X)$. One observes that the untwisted system
$(\A,\theta,X)$ always has an obvious covariant representation
$(\H,r,T)$, with $\H:=L^2(X)$ (with the Haar measure), $r(a)\equiv
a(Q)=$ multiplication with $a$ and $[T(y)u](x):=[U(-y)u](x)=
u(x+y)$. For $X=\R^N$ it coincides with the ``untwisted'' Schr\"odinger
representation introduced in Subsection {\bf \ref{schrorep}}. Let us now
choose $\lambda\in C^1\big(X;\CC(X;\T)\big)$ such that
$\delta^1(\lambda)=\omega$ (identity in $Z^2\big(X;\CC(X;\T)\big)$).
We set $T^\lambda(y):=r\big(\lambda(y)\big) T(y)$. Explicitly, for any
$x\in X$ and $u\in\H$, $\left[T^\lambda(y)u\right](x) = [\lambda(y)](x)u(x+y)
\equiv \lambda(x;y)u(x+y)$. Let us already
mention that the point (b) in the next proposition is at the root of
gauge invariance for magnetic pseudodifferential operators.

\begin{proposition}\label{corez}
\begin{itemize}
\item[{\rm (a)}] $(\H,r,T^\lambda)$ is a covariant representation of
$(\A,\theta,\omega,X)$,
\item[{\rm (b)}] If $\mu$ is another element of $C^1\big(X;\CC(X;\T)\big)$
such that $\delta^1(\mu)=\omega$, then there exists $c \in  \CC(X;\T)$
such that $\mu(x)=\frac{\theta_x(c)}{c}
\lambda(x)$, $\forall x\in X$. Moreover, $T^\mu(x)=
r(c^{-1})\;\!T^\lambda(x)\;\!r(c)$ for all $x\in X$.
\end{itemize}
\end{proposition}

\begin{proof}
The proof of the first statement consists in trivial
verifications. For the second statement, one first notes that
$\frac{\mu}{\lambda}$ belongs to $\ker(\delta^1)=Z^1
\big(X;\CC(X;\T)\big)$. Since this set is equal to
$B^1\big(X;\CC(X;\T)\big)$ by Lemma \ref{triv}, there exists $c \in
C^0\big(X; \CC(X;\T)\big) \equiv \CC(X;\T)$ satisfying $\mu(x)=
\frac{\theta_x(c)}{c} \lambda(x)$, $\forall x\in X$. The last
claim of the proposition follows from $r[\theta_x(c)]T(x)=T(x)r(c)$.
\end{proof}

We call $(\H,r,T^\lambda)$ {\it the Schr\"odinger covariant
representation associated with the $1$-cochain} $\lambda$. Let us now
recall the detailed form of the composition laws on $L^1(X;\A)$. For
simplicity we shall use notations as $\phi(x;y)$ for $[\phi(y)](x)$
and $\omega(x;y,z)$ for $[\omega(y,z)](x)$.
With these notations and for any $\phi,\psi \in L^1(X;\A)$,
the relations (\ref{produit}) and (\ref{invo}) read respectively
\begin{equation*}
(\phi\d^\omega_\tau \psi)(q;x)=\int_Xdy\;\!\phi\big(q + \tau(y-x);
y\big)\;\!\psi\big(q+(\mathfrak 1-\tau)y;x-y\big)\;\!\omega\big(q-\tau x;y,x-y\big)
\end{equation*}
and
\begin{equation*}
(\phi^{\d^\omega_\tau})(q;x)=\omega\big(q-\tau x;x,
-x\big)^{-1}\;\!\overline{\phi\big(q +(\mathfrak 1-2\tau)x;-x\big)},
\end{equation*}
where $x,y,q$ are elements of $X$.

Let us also denote for convenience by $\Rep^\lambda_\tau$ the
representation $r\rtimes_\tau T^\lambda$ in $L^2(X)$ of the twisted
crossed product $\A\rtimes^\omega_{\theta,\tau}X$.
Its explicit action on $L^1(X;\A)$ is given by
\begin{equation*}
\left[\left(\Rep^\lambda_\tau(\phi)\right)u\right](x)=
\int_X dy\;\!\phi(x+\tau y;y)\;\!\lambda(x;y)\;\!u(x+y)=
\int_X dy\;\!\phi\big((\mathfrak 1-\tau)x+\tau y;y-x\big)\;\!\lambda(x;y-x)
\;\!u(y).
\end{equation*}

We gather some important properties of $\Rep^\lambda_\tau$ in

\begin{proposition}\label{dada}
\begin{itemize}
\item[{\rm (a)}] In the setting of Proposition \ref{corez} (b), one has
$\;\Rep^\mu_\tau(\phi)=r(c^{-1})\Rep^\lambda_\tau(\phi)r(c)$.
\item[{\rm (b)}] $\Rep^\lambda_\tau[\CO(X)\rtimes_{\theta,\tau}^{\omega}X]
= \K\big(L^2(X)\big)$, the $C^*$-algebra of all compact operators in $L^2(X)$.
\item[{\rm (c)}] If $\;\CO(X)\subset\A$, then $\Rep^\lambda_\tau$
is irreducible.
\item[{\rm (d)}] $\Rep^\lambda_\tau$ is faithful.
\end{itemize}
\end{proposition}

\begin{proof}
(a) The proof of this statement consists in a simple verification.

(b) By Lemma \ref{triv}, $\omega$ belongs to $B^2\big(X;\CC(X;\T)\big)$,
i.e.~there exists $\lambda \in C^1\big(X;\CC(X;\T)\big)$
such that $\delta^1(\lambda) = \omega$. We may then consider the following
isomorphism
\begin{equation}\label{trv}
i^{\lambda^{-1}}_\tau : \left(L^1\big(X;\CC_0(X)\big), \d^1_{\0}, {}^{\d^1_{\0}}\right)
\to \left(L^1\big(X;\CC_0(X)\big), \d^{\omega}_{\tau},
{}^{\d^{\omega}_{\tau}}\right), \quad \left[i^{\lambda^{-1}}_\tau(\phi)\right](x)=
\theta_{-\tau x}\left[\lambda^{-1}(x)\;\!\phi(x)\right],
\end{equation}
that extends to an isomorphism between the non-twisted crossed
product $\CC_0(X)\rtimes_{\theta,\0}^{1}X$ and our $\CO(X)
\rtimes_{\theta,\tau}^{\omega}X$ (this is consistent with Remark \ref{remarca}).
One easily checks that $\Rep^{\lambda}_{\tau}\big[i^{\lambda^{-1}}_\tau(\phi)\big]
=  \int_X dx\;\!r[\phi(x)]T(x)$ for all $\phi$ in $\left(L^1(X;\A),
\d^1_{\0}, {}^{\d^1_{\0}}\right)$. But it is known that the image of
$\CO(X)\rtimes^{1}_{\theta,\0}X$ through the representation
$r\rtimes T\equiv\Rep^1_\0$ is equal to the algebra $\K\big(L^2(X)\big)$
of compact operators in $L^2(X)$, cf.~for example \cite{Georgescu1}.

(c) If $\CO(X) \subset \A$ then $\CO(X)\rtimes^\omega_{\theta,\tau}X$
can be identified to a $C^*$-subalgebra of $\A\rtimes^\omega_{\theta,\tau}X$
and the irreducibility of
$\Rep_{\tau}^{\lambda}\left(\A\rtimes_{\theta,\tau}^{\omega}X\right)$
follows from the irreducibility of $\K\big(L^2(X)\big)$, by (b).

(d) Let us recall from \cite{Raeburn1} the regular representation of the
twisted dynamical system $(\A,\theta,\omega,X)$. It consists in the triple
$(\H', r', T')$, where $\H'$ is the Hilbert space $L^2\big(X;L^2(X)\big)$, and
where the two maps act on $\xi \in \H'$ as
\begin{equation*}
\left[r'(a)\;\!\xi\right](x)=\theta_x(a)\;\!\xi(x) \quad \hbox{ and }
\left[T'(y)\;\!\xi\right](x)= \omega(x,y)\;\!\xi(x+y) \qquad \hbox{for all }
x, y \in X  \hbox{ and } a \in \A.
\end{equation*}
It follows by straightforward verifications that $(\H', r', T')$ is a
covariant representation.

Since $\H'$ is canonically isomorphic to $L^2(X \times X)$, let us set
$\xi(\cdot; x):= \xi(x)$ and introduce the unitary operator
$W^{\lambda}: L^2(X \times X) \to L^2(X \times X)$, $[W^{\lambda}\xi](x;y)
:=\lambda(x;y)\;\! \xi(x;x+y)$. Its adjoint is given by $[(W^{\lambda})^*\xi]
(x;y)= \lambda^{-1}(x;y-x)\;\!\xi(x;y-x)$.
Some easy calculations show then that $\left[(W^{\lambda})^*\;\!r'(a)\;\!
W^{\lambda}\;\!\xi\right](x;y)= a(y)\;\!\xi(x;y)$. Moreover, one has
\begin{equation*}
\left[(W^{\lambda})^*\;\!T'(z)\;\!W^{\lambda}\;\!\xi\right](x;y) =
\lambda^{-1}(x;y-x)\;\!\omega(x;y-x,z)\;\!\lambda(x;y-x+z)\;\!
\xi(x;y+z) = \lambda(y;z)\;\!\xi(x;y+z),
\end{equation*}
where we have used that $\omega = \delta^1(\lambda)$. Equivalently, one has
$(W^{\lambda})^*\;\!r'(a)\;\!W^{\lambda} = {\mathbf 1} \otimes
a(Q)$ and $(W^{\lambda})^*\;\!T'(z)\;\!W^{\lambda} = {\mathbf 1}
\otimes \lambda(Q;z) T(z) \equiv {\mathbf 1} \otimes T^{\lambda}(z)$ in
$L^2(X)\otimes L^2(X)$.
Thus the regular representation is unitarily equivalent to the
representation $(\H \otimes \H, {\mathbf 1}\otimes r, {\mathbf 1} \otimes
T^{\lambda})$. Since the regular representation induces a faithful
representation $r'\rtimes T'$ of $\A\rtimes_{\0}^{\omega}X$ in $\H'$,
cf.~Theorem 3.11 of \cite{Raeburn1} ($X$ is amenable, being abelian),
the Schr\"odinger representation induces faithful representations of
$\A\rtimes_{\theta,\tau}^{\omega}X$ in $\H$ for any $\tau\in\End(X)$.
\end{proof}

\begin{remark} {\rm If $\CC_0(X)$ is not contained in $\A$, then the
conclusion in (c) may fail. If, for example, $\A=\C$ (with the
trivial action) and $\omega=1$ then any translation $T(x)$ commutes with
$\Rep(\C\rtimes X)$, thus $\Rep(\C\rtimes X)$ is reducible by Schur's Lemma.}
\end{remark}

\begin{remark}
{\rm It is well-known that $\K\big(L^2(X)\big)$ admits a single class
of irreducible representations. Thus, by (c) and (d), all the
irreducible representations of $\CO(X)\rtimes_{\theta,\tau}^{\omega}X$
are unitarily equivalent to the Schr\"odinger representations
$\Rep^{\lambda}_{\tau}$ $\big($which form {\it always} a single class, by
Proposition \ref{dada} (a)$\big)$. This is by no means a general
property. Assume that $\A$ contains $\CC_0(X)$ and is unital. Then
its spectrum $\SA$ is a compactification of $X$. Since $X$ sits in
$\SA$ as a dense, $X$-invariant open set, there exist closed
invariant subsets of $\SA\setminus X$. Let $F$ be one of them; it
will be called {\it an asymptotic set}. For instance, one can ask it
to be minimal with respect to the properties above, i.e. it will be a
quasi-orbit disjoint of $X$; this will be assumed in the sequel. $\A$
being identified with $\CC(\SA)$, $\;\CC^F(\SA):=\{a\in\A\mid
a|_F=0\}$ is obviously an invariant ideal that we call $\A^F$. It is
easy to see that the multiplier algebra of $\A^F$ contains the
multiplier algebra of $\A$, so that, by restriction, the twisted
dynamical system $(\A^F,\theta,\omega,X)$ makes sense. The twisted
crossed product $\A^F\rtimes_{\theta,\tau}^{\omega}X$ may be
identified with an ideal of $\A\rtimes_{\theta,\tau}^{\omega}X$ and the
quotient to $\left(\A/\A^F\right)\rtimes_{\theta,\tau}^{\omega^F}X$.
To understand $\omega^F$, remark that $\A/\A^F$ is canonically isomorphic
to the $C^*$-algebra $\CC(F)$ of all continuous functions on $F$.
In this interpretation, for each $x,y\in X$, $\omega(x,y)\in\U(\A)$
first extends to $\SA$ and then is restricted to $F$, giving thus a
$2$-cocycle $\omega^F:X\times X\rightarrow \U[\CC(F)]=\CC(F;\T)$.
If we choose now an irreducible representation
$R^F$ of the quotient, one gets an irreducible representation of the
initial twisted crossed product just by composing with the quotient
map. This representation is no longer faithful, hence it cannot be
unitarily equivalent to the initial one.
We can choose for $R^F$ once again a Schr\"odinger-type representation,
but associated with the asymptotic twisted dynamical system
$(\CC(F),\theta,\omega^F,X)$.
To do this, we need to show that it is standard. This follows if we
interpret $\CC(F)$ as an $X$-algebra. This is done simply by choosing
a point $\nu_0$ on $F$ whose quasi-orbit is the entire $F$.
This will lead to an embedding of $\CC(F)$ into $\BC_u(X)$.
Thus the quotient is once again a standard twisted dynamical system,
but with a ``simpler'' $C^*$-algebra and a ``simpler'' $2$-cocycle
$\omega^F$. Then one can fix a pseudo-trivialization $\lambda^F$
of $\omega^F$ and thus we may take $R^F=\Rep^{\lambda^F}_\tau$.}
\end{remark}

\begin{remark}
{\rm Let us summary some facts obtained above, that will be relevant in
the next sections. We fix a standard twisted dynamical system
$(\A,\theta,\omega,X)$. The $C^*$-algebras
$\A\rtimes_{\theta,\tau}^\omega X$ are isomorphic to each other
for different elements $\tau \in \End(X)$. This also has a
cohomological nature, but in Subsection {\bf \ref{s3.2}} we shall be
in a better position to discuss this. So let us fix $\tau$. For some choices
of $\A$, $\A\rtimes_{\theta,\tau}^\omega X$ may be isomorphic to the
untwisted crossed product $\A\rtimes_{\theta,\tau}^1 X$. This happens
when $\omega$ is a 2-coboundary with respect to the unitary group of
$\A$; the typical case is $\A = \CO(X)$. Definitely, this does not
happen too often. In general $\omega$ can be
trivialized only by using a larger unitary group and this is not enough
to conclude that it can be removed from
$\A\rtimes_{\theta,\tau}^\omega X$ by an isomorphism. On the other
hand, for fixed $\omega$, one constructs a family of Schr\"odinger
representations of $\A\rtimes_{\theta,\tau}^\omega X$ indexed by all
the 1-cochains which define the pseudo-trivializable $\omega$ with
respect to the larger $X$-module $\CC(X;\T)$. This is possible since the
representation $r$ of the algebra $\A$ is a restriction of a much
larger representations. The Schr\"odinger representations assigned to
two 1-cochains giving he same $\omega$ are unitary equivalent. This happens
since the 1-cochains are cohomologous, once again with respect to the big
$X$-module $\CC(X;\T)$. This phenomenon is a general instance of what is
called ``gauge covariance'' for magnetic algebras and representations.}
\end{remark}

\section{The pseudodifferential calculus associated with a twisted
dynamical system}\label{s3}

We have introduced a class of $C^*$-algebras,
called standard twisted crossed products, as well as their family
of Schr\"odinger representations, defined by pseudo-trivializations
of the $2$-cocycle $\omega$.
In the first subsection, by a partial Fourier transformation, we
shall get from these data a sort of pseudo\-differential calculus.
Let us denote by $X^\sharp$ the dual group of $X$.
Then certain classes of functions on $X\times X^\sharp$ will be
organised in $C^*$-algebras with some natural involution and a
product involving $\omega$ and generalizing the well-known Moyal
product appearing in Quantum Mechanics. The composition between
the partial Fourier transformation and the Schr\"odinger representation
will lead to a rule of assigning operators to symbols
belonging to these $C^*$-algebras. This will be a generalization
of the pseudodifferential (in particular of the Weyl) rule
valid for $X=\R^N$ in the absence of any $2$-cocycle.

The axioms of a covariant representation are a sort of a priori
commutation relations. We can reinterpret them in the form of a Weyl
system - a family of unitary operators satisfying a relation which
generalizes that of a projective representation. This Weyl system will
be introduced and studied in the second subsection. We also show that
the pseudodifferential prescription may be considered as a functional
calculus associated with this Weyl system, by mimicking an approach
that is standard in the commutative case.

\subsection{Generalized pseudodifferential algebras and
operators}\label{s3.1}

Let $X^{\sharp}$ be the dual group of $X$, i.e.~the set of all
continuous morphisms (characters) $\;\chi:X \to \T$. Endowed with the
composition law $\;(\chi\cdot\kappa)(x):=\chi(x)\kappa(x)$, $x\in X$
and with the topology of uniform convergence on compact subsets of $X$,
$\;X^{\sharp}$ is a second-countable locally compact abelian group. The Haar
measures on $X$ and $X^{\sharp}$ will be normalized in such a way
that the Fourier transformations
\begin{equation*}
\FX : L^1(X) \to \CO(X^{\sharp}), \quad \left(\FX \; b\right)(\chi)=
\int_X dx\;\! \overline{\chi(x)}\;\!b(x)
\end{equation*}
and
\begin{equation*}
\FXb : L^1(X) \to \CO(X^{\sharp}), \quad \left(\FXb \; b\right)(\chi)=
\int_X dx\;\! \chi(x)\;\!b(x)
\end{equation*}
induce unitary maps from $L^2(X)$ to $L^2(X^{\sharp})$. The inverses
of these maps act on $L^2(X^\sharp)\cap L^1(X^\sharp)$ as
$\left(\FXSb \;\! c\right)(x)= \int_{X^\sharp} d\chi\;\! \chi(x)\;\! c(\chi)$
and $\left(\FXS \;\! c\right)(x)= \int_{X^\sharp} d\chi\;\! \overline{\chi(x)}
\;\! c(\chi)$.

Let us now consider the twisted dynamical system
$(\A,\theta,\omega,X)$. We define the mapping
$\mathbf 1\otimes\FXb :L^1(X;\A)\to\CC_0(X^{\sharp};\A)$ by
$\left[\left({\mathbf 1}\otimes\FXb \right)(\phi)\right]
(\chi)= \int_X dx\;\! \chi(x)\;\! \phi(x)$ (equality in
$\A$). We recall that $\A\odot L^1(X)$ is a dense subspace of $L^1(X;\A)$
and observe that $\left({\mathbf 1}\otimes\FXb \right)(a\otimes b)=
a\otimes \left(\FXb\;\! b \right)$. Let us now also fix an element
$\tau\in\End(X)$. We transport  all the structure of the Banach $^*$-algebra
$(L^1(X;\A),\diamond^\omega_\tau,^{\diamond^\omega_\tau},\n\cdot\n)$ to the
corresponding subset of $\CC_0(X^{\sharp};\A)$
via ${\mathbf 1}\otimes\FXb $. The space $\left({\mathbf 1}\otimes\FXb \right)
L^1(X;\A)$ will also be a Banach $^*$-algebra with a composition
law $\circ^\omega_\tau$, an involution $^{\circ^\omega_\tau}$ and the
norm $\n(\mathbf 1\otimes\FXb^{-1})\cdot\n$.
Its envelopping $C^*$-algebra will be denoted by
$\mathfrak C^\omega_{\A,\tau}$. The map ${\mathbf 1}\otimes\FXb$
extends canonically to an isomorphism between
$\A\rtimes_{\theta,\tau}^\omega X$ and $\mathfrak C^\omega_{\A,\tau}$.
We remark that $\left({\mathbf 1}\otimes\FXb\right)\left[\A\odot
L^1(X)\right]$ is already not very explicit, since one has no direct
characterization of the space $\FXb\left[L^1(X)\right]$.
Concerning $\mathfrak C^\omega_{\A,\tau}$, we do not even know if it
consists entirely of $\A$-valued distributions on $X^\sharp$
(whenever this makes sense). However, usually one can work efficiently
on suitable dense subsets.

We deduce now the explicit form of the composition law and of the involution.
Let us denote simply ${\mathbf 1}\otimes\FXb$ by $\mathfrak F$. One gets for any
$f,g\in\mathfrak F L^1(X;\A)$
\begin{eqnarray*}
& (f\circ^\omega_\tau g)(q,\chi):=\left(\mathfrak F\left[(\mathfrak F^{-1}f)
\diamond^{\omega}_\tau (\mathfrak F^{-1}g)\right]\right)(q,\chi)= & \\
& \int_{X}dx\int_{X}dy\int_{X^\sharp}d\kappa\int_{X^\sharp}d\gamma\ \!
\chi(x)\;\!\overline{\kappa(y)}\;\!\overline{\gamma(x-y)}\;\!f(q+\tau(y-x),
\kappa)\;\!g(q+(\mathfrak 1-\tau)y,\gamma)\;\!\omega(q-\tau x;y,x-y) &
\end{eqnarray*}
and
\begin{equation*}
\left(f^{\circ^\omega_\tau}\right)(q,\chi):=\left(\mathfrak F\left[\left(
\mathfrak F^{-1}f\right)^{\diamond^{\omega}_\tau}\right]\right)(q,\chi)=
\int_{X}dx\int_{X^\sharp}d\kappa\;\!\left(\chi\cdot\kappa^{-1}\right)(x)\;\!
\omega(q-\tau x;x,-x)^{-1}\;\!\overline{f(q+(1-2\tau)x,\kappa)}.
\end{equation*}
Both expressions make sense as iterated integrals; under more stringent
conditions on $f$ and $g$, the integrals will be absolutely convergent.

The constructions and formulae above can be given (with some slight
adaptations) for general abelian twisted dynamical system. We ask now
that our twisted dynamical system be standard, which makes $\omega$
pseudo-trivial. For any continuous function $\lambda:X\to\CC(X;\T)$
such that $\delta^1(\lambda)= \omega$, the corresponding Schr\"odinger
covariant representation $(\H,r,T^{\lambda})$ gives rise to the
Schr\"odinger representation of $\A\rtimes_{\theta,\tau}^\omega X$ that
we denoted by $\Rep^\lambda_\tau$. We get a representation of
$\mathfrak C^\omega_{\A,\tau}$ just by composing with $\mathfrak F^{-1}$;
it will be denoted by $\Op^\lambda_\tau$. By simple calculations one obtains

\begin{proposition}
\begin{enumerate}
\item[{\rm (a)}] The representation $\Op^\lambda_\tau:=\Rep^\lambda_\tau
\circ\mathfrak F^{-1}:\mathfrak C^\omega_{\A,\tau}\rightarrow \B(\H)$ is
faithful and acts on $f\in \mathfrak FL^1(X;\A)$ by the formula
\begin{equation}\label{formula}
\left[\Op^\lambda_\tau(f)u\right](x)=\int_Xdy\;\!\int_{X^\sharp}d\chi\ \!
\chi(x-y)\;\!\lambda(x;y-x)\;\!f[(\mathfrak 1-\tau)x+\tau y,\chi]\;\!u(y),
\ \ u\in\H,\ \ x\in X,
\end{equation}
where the right-hand side is viewed as an iterated integral.
\item[{\rm (b)}] If $\mu\in C^1\big(X;\CC(X;\T)\big)$ is another $1$-cochain,
giving a second pseudo-trivialization of the $2$-cocycle $\omega$,
then $\mu=\delta^0(c)\lambda$ for some $c\in\CC(X;\T)$ and
$\Op^\lambda_\tau$, $\Op^\mu_\tau$ are unitarily equivalent:
\begin{equation}\label{unechiv}
r(c^{-1})\;\!\Op^\lambda_\tau(f)\;\!r(c)=\Op^\mu_\tau(f),\ \ \
\forall f\in\mathfrak C^\omega_{\A,\tau}.
\end{equation}
\end{enumerate}
\end{proposition}

In (\ref{formula}) the integral is absolutely convergent under various
assumptions on $f$ and $u$, for instance if $u\in L^1(X)\cap L^2(X)$ and
$f$ is of class $L^1$ in $\chi$. One could dare and call (\ref{unechiv})
{\it the gauge-covariance of the generalized pseudodifferential calculus}.

For different $\tau$'s, the $C^*$-algebras $\mathfrak C^\omega_{\theta,\tau}$
are isomorphic. If $\tau,\tau'\in\End(X)$, then $\mathfrak m_{\tau,\tau'}
:=\mathfrak F\circ m_{\tau,\tau'}\mathfrak F^{-1}$ defines an isomorphism
$\mathfrak C^\omega_{\theta,\tau'}\cong\mathfrak C^\omega_{\theta,\tau}$.
We recall that $\left(m_{\tau,\tau'}\phi\right)(q;x)=\phi(q+(\tau'-\tau)x;x)$,
$\forall x,q\in X$, $\forall \phi\in L^1(X;\A)$.
This isomorphism is constructed in order to satisfy $\Op^\lambda_{\tau'}=
\Op^\lambda_\tau\circ\mathfrak m_{\tau,\tau'}$ and thus gives the
transformation of the $\tau$-symbol of a generalized pseudodifferential
operator into its $\tau'$-symbol.

We support the assertion that the choice of the parameter $\tau$ is a matter
of ordering only by a weak example. Let us assume that the $X$-algebra $\A$
is unital. Then the element $f=1\otimes b$ is in
$\mathfrak C^\omega_{\theta,\tau}$ for any $b:X^\sharp\rightarrow \C$ with
$\FXS \;\! b\in L^1(X)$. The operator
$\Op^\lambda_\tau(1\otimes b)$ does not depend on $\tau$. We denote it by
$\mathfrak{op}^\lambda(b)$; its action on $u\in\H$ is given by
$$ \left[\mathfrak{op}^\lambda(b)u\right](x)=\int_Xdy\;\lambda(x;y-x)
\left(\FXS \;\! b\right)(y-x)u(y).$$
Let us now consider an arbitrary element $a\in\A$.
Simple calculations for $\tau=\mathfrak 0$ and $\tau=\mathfrak 1$ show that
$\Op^\lambda_\mathfrak 0(a\otimes b)=r(a)\mathfrak{op}^\lambda(b)$ and
$\Op^\lambda_\mathfrak 1(a\otimes b)=\mathfrak{op}^\lambda(b)r(a)$.
We point out that $\;b\to\mathfrak{op}^\lambda(b)$ is not a closed functional
calculus: $(1\otimes b_1)\circ^\omega_\tau(1\otimes b_2)$ is in general a
function depending on both variables, hence one cannot write
$\mathfrak{op}^\lambda(b_1)\mathfrak{op}^\lambda(b_2)=\mathfrak{op}^\lambda(b)$
for some function $b$ defined on $X^\sharp$. It is not difficult to extend
the morphism $f\to\Op^\lambda_\tau(f)$ to include elements $f=a\otimes b$
with $a\in\A$ and $b:X^\sharp\rightarrow\mathbb C$ being the Fourier transform
of some bounded measure on $X$. Then we see that $\Op^\lambda_\tau(a\otimes
1)=r(a)$ for all $\lambda$ and $\tau$.

\subsection{Generalized Weyl systems and the functional
calculus}\label{s3.2}

From now on the product group $X\times X^\sharp$ will be denoted simply by
$\Xi$; it is locally compact, second-countable and abelian. We shall rephrase
the relations verified by the Schr\" odinger covariant representation
$(r,\H,T^\lambda)$ of the standard twisted dynamical system
$(\A,\theta,\omega,X)$, insisting on the role played by $\Xi$. In this way
we generalize the Weyl system of {\bf \ref{sub11}}.

Remark first that $r$ is the restriction to $\A$ of a representation
(also denoted by $r$) of $\B\CC(X)$. On the other hand, the dual group
$X^\sharp$ is naturally a subset of $\B\CC(X)$ (any character is a continuous
function on $X$ with values in the bounded subset $\T$ of $\C$). Thus we can
consider the map $V:X^\sharp\rightarrow\U(\H)$, $V(\chi):=r(\chi)^*=$operator
of multiplication by $\overline{\chi}$ in $\H=L^2(X)$. Obviously $V$ is a
unitary group representation, the same appearing in Subsection {\bf
\ref{sub11}} for the particular case $X=\R^N$. We set also
$U^\lambda(x):=T^\lambda(x)^*$ for any $x\in X$. By putting
$a=\overline{\chi}$ in the covariance relation
$\;T^\lambda(x)r(a)T^\lambda(x)^*=r[\theta_x(a)]$ and using
$\theta_x(\overline{\chi})=\overline{\chi(x)}\;\overline{\chi}$, one gets
readily
\begin{equation}
U^\lambda(x)V(\chi)=\chi(x)V(\chi)U^\lambda(x),\ \ \forall (x,\chi)\in\Xi,
\end{equation}
which is a generalization of the Weyl form (\ref{WCCR}) of the canonical
commutation relations.

Let us also fix an endomorphism $\tau$ of the group $X$. By generalizing
(\ref{wey}) one sets for all $\xi=(y,\chi)\in\Xi$
\begin{equation*}
W_{\tau}^{\lambda}(y,\chi):= \chi[(\mathfrak 1-\tau)y]\;\!U^{\lambda}(y)^*
\;\!V(\chi)=\chi[-\tau y]\;\!V(\chi)\;\!U^{\lambda}(y)^*.
\end{equation*}
Explicitly, one has for $u\in\H$
\begin{equation}\label{explic}
\left[W_{\tau}^{\lambda}(y,\chi)u\right](x)=\chi[-x-\tau y]\;\!\lambda(x;y)
\;\!u(x+y).
\end{equation}

\begin{definition}
{\rm The family of unitary operators $\{W^\lambda_\tau(\xi)\}_{\xi\in\Xi}$
is called} the Weyl system associated with the pseudo-trivialization
$\lambda$ and the endomorphism $\tau$.
\end{definition}

These operators satisfy for all $\xi=(x,\chi),\eta=(y,\kappa)\in\Xi$
the relations
\begin{equation}\label{WS}
W_{\tau}^{\lambda}(x,\chi)\;\!W_{\tau}^{\lambda}(y,\kappa)=
r\{\chi[\tau y]\;\!\kappa[(\tau-1) x]\;\!\omega(x,y)\}\;\!W_{\tau}^{\lambda}
(x+y,\chi\cdot\kappa).
\end{equation}
In fact this is part of a more comprehensive assertion:

\begin{proposition}
\begin{enumerate}
\item[{\rm (a)}] If $(r,\H,T^\lambda)$ is a covariant representation of
$\;(\A,\theta,\omega,X)$, then $(r,\H,W^\lambda_\tau)$ is a covariant
representation of the twisted dynamical system
$\;(\A,\Theta,\Omega_\tau,\Xi)$, where $\Xi=X\times X^\sharp$,
$\;\left[\Theta_{(x,\chi)}(a)\right](y)=\left[\theta_{x}(a)\right](y)
=a(y+x)$ and $\;\Omega_\tau[(x,\chi),(y,\kappa)]:=\chi[\tau y]
\;\!\kappa[(\tau-1) x]\;\omega(x,y)$.
\item[{\rm (b)}] If $\mu$ is another element of $C^1\big(X;\CC(X,\T)\big)$
such that $\delta^1(\mu)=\omega$, then there exists $c \in  \CC(X;\T)$
such that $W^{\mu}_{\tau}(\xi)= r(c^{-1})\;\!W^{\lambda}_{\tau}(\xi)\;\!
r(c)$ for all $\xi\in\Xi$.
\item[{\rm (c)}] For $\tau,\tau'\in\End(X)$, the $2$-cocycles $\Omega_\tau$ and
$\Omega_{\tau'}$ on $\Xi$ are cohomologous and the corresponding Weyl
systems are connected by $W^{\lambda}_{\tau'}(x,\chi)=\chi[(\tau - \tau')x]
\;\!W^{\lambda}_{\tau}(x,\chi)$ for all $x$ and $\chi$ .
\end{enumerate}
\end{proposition}

\begin{proof}
(a) Simple verifications.

(b) This follows from Proposition \ref{corez} (b) or by direct calculation.

(c) One finds immediately that $\Omega_{\tau'}[(x,\chi),(y,\kappa)]=
\chi[(\tau'-\tau)y]\kappa[(\tau'-\tau)x]\Omega_\tau[(x,\chi),(y,\kappa)]$,
which can be written $\Omega_{\tau'}=\delta^1(\Lambda_{\tau,\tau'})
\Omega_\tau$ for $\Lambda_{\tau,\tau'}(x,\chi)=\chi[(\tau-\tau')x]$.
The relation between $W^{\lambda}_{\tau}$ and $W^{\lambda}_{\tau'}$ follows
then from (\ref{WS}) or is deduced directly from the explicit formula (\ref{explic}).
\end{proof}

The inflated twisted dynamical system $(\A,\Theta,\Omega_\tau,\Xi)$ may be
used to construct twisted crossed product $C^*$-algebras. We do not pursue
this here. The point (c) shows that the correlation between the structures
defined by different $\tau$'s is once again a matter of cohomology.

Let us denote by $\F_\Xi$ the ``symplectic'' Fourier transformation defined
on $L^1(\Xi)$ by
\begin{equation*}
\left(\mathcal F_\Xi g\right)(x,\chi):=\int_X\int_{X^\sharp}dy\;d\kappa\;
\chi(y)\overline{\kappa(x)}g(y,\kappa),
\end{equation*}
which can be expressed as $\mathcal F_\Xi=\mathcal I\circ\left(\FXb \otimes
\FXS \right)$, with $(\mathcal Ih)(x,\chi) :=h(\chi,x)$.
One easily checks that $\F_\Xi^{-1} = \F_\Xi$.
It is natural to define for $f\in\mathcal F_\Xi L^1(\Xi)$
\begin{equation}\label{strateg}
\widetilde{\Op}^{\lambda}_\tau(f)=\int_\Xi d\xi\;(\F_{\Xi}^{-1}f)(\xi)
W_{\tau}^{\lambda}(\xi).
\end{equation}
This is intended to be a sort of functional calculus generalizing
(\ref{incep}), the idea being to mimick once again a formula that works
well in the simple, commutative case. A certain convention used in
constructing our $W^\lambda_\tau$ asks for the ``symplectic'' Fourier
transformation.

The next result will show that we have already constructed this functional
calculus. We need to take into account the algebraic tensor product
(finite linear combinations of elementary tensors) $\mathfrak L:=
\FXS L^1(X^\sharp)\odot\FXb L^1(X)$.
It can be naturally viewed as a subspace of $\CC_0(X)\odot\CC_0(X^\sharp)$.

It is simple to check that $\mathfrak L$ is a subspace of
$\mathcal F_\Xi L^1(\Xi)$ (on which $\widetilde{\Op}^{\lambda}_\tau$ is
defined). If $\CC_0(X) \subset \A$, then $\mathfrak L$ is also a subspace
of $\A\odot\FXb L^1(X)\subset \mathfrak F L^1(X;\A)$
(on which $\Op^{\lambda}_\tau$ is defined). We show that
$\widetilde{\Op}^{\lambda}_\tau(f)=\Op^{\lambda}_\tau(f)$ for the elementary
vector $f=\left(\FXS \;\! a\right)\otimes\left(\FXb \;\! b\right)$, with
$a\in L^1(X^\sharp)$ and $b\in L^1(X)$. Note first that
\begin{equation*}
\mathcal F_\Xi^{-1}f=\left(\FXb \otimes \FXS \right)^{-1}
\left\{\mathcal I^{-1}\left[\left(\FXS \;\! a\right)
\otimes\left(\FXb \;\!b\right)\right]\right\}=
\left(\FXS \otimes \FXb \right)\left\{\left(\FXb \;\! b\right)\otimes
\left(\FXS \;\!a\right)\right\}=b\otimes a.
\end{equation*}
Then
\begin{equation*}
\widetilde{\Op}^{\lambda}_\tau(f)=\int_\Xi d\xi\;(b\otimes a)(\xi)
W^\lambda_\tau(\xi)=\int_Xdy\;r\left[b(y)\int_{X^\sharp}d\chi\;a(\chi)
\chi(-\tau y)\overline{\chi}\right]T^\lambda(y)=
\end{equation*}
\begin{equation*}
=\int_Xdy\;r\left[\theta_{\tau y}\big([\FXS \;\! a\otimes b](y)
\big)\right]T^\lambda(y)=\Op^{\lambda}_\tau(f).
\end{equation*}
Thus we have proved
\begin{proposition}\label{minunea}
Assume that the abelian $X$-algebra $\A$ contains $\;\CC_0(X)$.
Then both $\Op^{\lambda}_\tau$ and $\widetilde{\Op}^{\lambda}_\tau$ are
well-defined on $\mathfrak L$ and they coincide on this set.
\end{proposition}
We regard (\ref{formula}) and (\ref{strateg}) as special instances of the
same object that makes sense for more general classes of symbols $f$,
maybe in a weaker sense.

\subsection{Extensions}\label{s3.3}

One would like to extend the composition laws $\,\diamond^\omega_\tau $
and $\,\circ^\omega_\tau$ and the representations $\,\Rep^\lambda_\tau$
and $\,\Op^\lambda_\tau$ to more general symbols. We shall indicate only
the extension results making use of multiplier algebras. As a rule, the
extensions will be denoted by the same letters as before.

The general theory says that any $C^*$-algebra $\mathfrak C$ is embedded
as an essential ideal in the (maximal) multiplier $C^*$-algebra
$\mathfrak M(\mathfrak C)$ and that any non-degenerate representation of
$\mathfrak C$ extends to a representation of $\mathfrak M(\mathfrak C)$.
This should be applied respectively to the $C^*$-algebra
$\A\rtimes^\omega_{\theta,\tau} X$ with the representation
$\,\Rep^\lambda_\tau$ and to the $C^*$-algebra $\mathfrak C^\omega_{\A,\tau}$
with the representation $\,\Op^\lambda_\tau$. We shall spell out only the
case $\A=\CC_0(X)$, that has some specific features.

Let us set $\mathcal N_{\diamond^\omega_{\tau}}:=
\mathfrak M\left[\CC_0(X)\rtimes^\omega_{\theta,\tau} X\right]$ and
$\mathcal N_{\circ^\omega_\tau}:=\mathfrak M\left[
\mathfrak C^\omega_{\CC_0(X),\tau}\right]$. The partial Fourier transform
$\mathfrak F$ extends to an isomorphism between these two $C^*$-algebras.
By (b) and (d) of Proposition \ref{dada}, $\,\Rep^\lambda_\tau$
defines an isomorphism between $\CC_0(X)\rtimes^\omega_{\A,\tau} X$ and the
ideal $\K(\H)$ of all compact operators in $\H=L^2(X)$.
So it extends to an isomorphism between the corresponding multiplier
algebras. But the multiplier algebra of $\K(\H)$ is the entire $\B(\H)$.
Concluding, the $C^*$-algebras $\mathcal N_{\diamond^\omega_\tau}$ and
$\mathcal N_{\circ^\omega_\tau}$ are both represented faithfully and
surjectively on $\B(\H)$ respectively by $\,\Rep^\lambda_\tau$ and
$\Op^\lambda_\tau$, the representations being connected to each other
by the extension of the isomorphism $\mathfrak F$.

Now, of course, $\mathcal N_{\diamond^\omega_\tau}$ and
$\mathcal N_{\circ^\omega_\tau}$ are defined only in a very implicit way.
Even in simple case ($X=\mathbb R^N$, $\omega=1$) one only knows that certain
classes of symbols belong to them. The Calder\'on-Vaillancourt Theorem
(cf.~\cite{Folland}) is such a statement and a generalization to the case
of a non-trivial $\omega$ would be interesting.
For our general situation we shall give only rather
simple results. The strategy is to extend explicitly $\,\Op^\lambda_\tau$ or
$\,\Rep^\lambda_\tau$ to the desired space of functions or measures,
by imposing that the corresponding operators are bounded.

\begin{proposition}
The space $\mathcal F_{\Xi}\mathbb M(\Xi)$ of all (symplectic) Fourier
transforms of bounded, complex measures is contained in
$\mathcal N_{\circ^\omega_\tau}$.
\end{proposition}

\begin{proof}
For $F\in\mathcal F_\Xi\mathbb M(\Xi)$, one defines $\,\Op^\lambda_\tau(F)
=\int_\Xi\left(\mathcal F_{\Xi}^{-1}F\right)(d\xi)W^\lambda_\tau(\xi)$ in a
weak, dual sense: for $u,v\in\H$, $\left<v,\Op^\lambda_\tau(F)u\right>$
is obtained by applying the bounded complex measure $\mathcal F_{\Xi}^{-1}F$
to the bounded continuous function $\left<v,W^\lambda_\tau(\cdot)u\right>$.
This defines bounded operators.
\end{proof}

Let us consider ``the exponential functions'' $\{F_{\xi_0}\}_{\xi_0\in\Xi}$,
where for $\xi_0=(x_0,\chi_0)\in\Xi$ one sets $F_{\xi_0}(x,\chi)=
\chi(x_0)\overline{\chi_0(x)}$. They are symplectic Fourier transforms of
Dirac measures, $F_{\xi_0}=\mathcal F_\Xi\delta_{\xi_0}$, thus elements of
$\mathcal{N}_{\circ^\omega_\tau}$. The Weyl system is obtained by applying
$\Op^\lambda_\tau$ to them:
$\Op^\lambda_\tau(F_{\xi_0})=W^\lambda_\tau(\xi_0)$, $\forall \xi_0\in\Xi$.
A simple calculation shows that linear combinations of functions $F_{\xi_0}$
do not form an algebra.

\subsection{Example: $X=\mathbb Z^2$}

The case $X=\Z^2$ leads, under some simplifying assumptions, to intensively
studied objects as the rotation algebra and almost Mathieu operators.
Our aim is to show that these objects emerge naturally and to allow
comparisons with Section \ref{s4}, in which $X=\mathbb R^N$.

We may work with any translational invariant $C^*$-subalgebra $\A$ of the
$C^*$-algebra of bounded complex functions on $\Z^2$. In most of our arguments
we shall take $\A=\mathbb C$ and this allows only the trivial action
$\theta_x=id$, $\forall x \in \mathbb Z^2$. Then one deals with $2$-cocycles
$\omega:\Z^2\times\Z^2\rightarrow \mathbb{T}\subset \mathbb{C}$. It follows
from Prop.~6.2 of \cite{Guichardet} that any such $2$-cocycle is
cohomologous with one of the form  $\omega^B(x,y)=\exp[-iB(x,y)]$, where
$B:\Z^2\times\Z^2\rightarrow\mathbb{R}$ is an antisymmetric biadditive map
(with a bit of imagination one could call it {\it a constant magnetic field
on the lattice}). To make the connection with standard notations, one sets
for some $\alpha\in\mathbb{R}$
\begin{equation}\label{alpha}
B(x,y)=\pi\alpha(x_1y_2-x_2y_1)\equiv\pi\alpha x\wedge y.
\end{equation}
and write $\omega_\alpha$ instead of $\omega^B$. Note that
$\omega_\alpha(x,-x)=0$, $\forall x\in \mathbb Z^2$.

Our twisted dynamical system is $(\mathbb{C},\text{\sl id}, \omega_\alpha ,
\Z^2)$. The space $L^1(\Z^2;\mathbb{C})\equiv L^1(\Z^2)$, endowed with the
structure given by
$$
(\varphi\diamond\psi)(x)=\sum_{y\in\Z^2}\omega_\alpha(y,x)\varphi(y)\psi(x-y)=
\sum_{y\in\Z^2}\exp\{-i\pi\alpha (y\wedge x)\}\varphi(y)\psi(x-y),
$$
$$
\varphi^\diamond(x)=\overline{\varphi(-x)},\qquad\n\varphi\n=\sum_{y\in\Z^2}
|\varphi(y)|
$$
is a Banach $^*$-algebra (the action $\theta$ being trivial, $\tau$ plays
no role at all here). We denote by $C^*_{\omega_\alpha}(\mathbb Z^2)\equiv
C^*_{\alpha}(\mathbb Z^2):=\mathbb{C}\rtimes^{\omega_\alpha}_{\text{\sl id}}
\Z^2$ the associated twisted crossed product. It is the twisted group
$C^*$-algebra associated with the group $\mathbb Z^2$ and the multiplier
$\omega_\alpha$. Traditionally it is called {\it the rotation algebra assigned
to the real number $\alpha$} and the usual notation is something like
$\mathfrak A_\alpha$. It can also be defined as the untwisted crossed product
of $\mathcal C(\mathbb T)$ by a suitable action of $\mathbb Z$.
For $\alpha=0$ we get the group $C^*$-algebra of $\Z^2$;
its spectrum is the $2$-torus $\mathbb T^2$. For $\alpha\ne 0$ one
obtains the so-called \textit{noncommutative tori of dimension 2}. Similar
objects may be defined for higher dimensions $N$ (i.e. for $X=\Z^N$),
cf.~\cite{Rieffel2}. In these cases the relevant input is an $N\times N$ real,
antisymmetric matrix $(\alpha_{jk})_{j,k=1,\dots,N}$.

The group $\mathbb Z^2$ being discrete, the $C^*$-algebra
$C^*_{\alpha}(\mathbb Z^2)$ is generated by the elements
$\{\delta_x\}_{x\in\Z^2}$ with $\delta_x(y):=0$ for $y\ne x$ and
$\delta_x(x):=1$, and is unital, with $\mathbf{1}=\delta_0$.
These generators satisfy the relations: $\delta_x\diamond\delta_y=
\omega_\alpha(x,y)\delta_{x+y}$, $\delta_x^\diamond=\delta_{-x}$.
In fact, since $\Z^2$ is generated as a group by the elements $(1,0),(0,1)$,
we recover easily the usual definition of the rotation algebra as the
universal $C^*$-algebra generated by two unitary elements $u:=\delta_{(1,0)}$
and $v:=\delta_{(0,1)}$ satisfying $u\diamond v=e^{2\pi i\alpha}v\diamond u$
(cf.~\cite{Rieffel1}, \cite{Bellissard1}, \cite{Boca} for instance).

One finds inside the rotation algebra elements with interesting spectral
properties (see for instance \cite {Bellissard2}, \cite {Boca}, \cite{Shubin2}
and references therein). Let $\nu$, $\mu$ be real numbers and $u$, $v$ the
elements introduced above. Then
$$
h(\alpha,\nu,\mu):=u+u^*+\nu\left(e^{2\pi i\mu}v+e^{-2\pi i\mu}v^*\right)
\in L^1(\mathbb Z^2)\subset C^*_\alpha(\mathbb Z^2)
$$
is called {\it the almost Mathieu Hamiltonian}, and the simplified version
$h(\alpha):=h(\alpha,1,0)$ is called {\it the Harper Hamiltonian}.

The example of {\bf \ref{sstand}} shows that for $\alpha\ne 0$ the
$2$-cocycles we consider are not trivial. But since our twisted dynamical
system is standard, by Proposition \ref{Couchepin}, all $\omega_\alpha$'s
are pseudo-trivial, i.e.~they are coboundaries with respect to the larger
Polish module $\mathcal C(\mathbb Z^2;\mathbb T)$.
Thus we can find a function $\lambda\in C^1\big(\mathbb Z^2;\mathcal C
(\mathbb Z^2;\mathbb T)\big)$ such that $\delta^1(\lambda)=\omega_\alpha$.
In fact one may take $[\lambda(y)](q):=\omega_\alpha(q,y)$ $\big($this is
exactly the choice (\ref{psetri})$\big)$. We set $\H=L^2(\mathbb Z^2)$
(with the counting measure) and consider the covariant representation
$(\H,r,T^\lambda)$ given for $u \in \H$ by:
 $$
r:\mathbb{C}\rightarrow \mathcal B(\H),\ \ \  r(\nu)u:=\nu u,
$$
$$
T^\lambda:\Z^2\rightarrow\mathcal{U}(\mathcal{H}),\qquad
[T^\lambda(y)u](x) :=\omega_\alpha(x,y)u(y+x).
$$
Then we get the associated representation $\mathfrak{Rep}^\lambda:=
r\rtimes T^\lambda:L^1(\Z^2)\rightarrow \mathcal B(\H)$
$$
\left[\mathfrak{Rep}^\lambda(\varphi) u\right](x):=\sum_{y\in\Z^2}
\omega_\alpha(x,y)\varphi(y)u(x+y),
$$
which extends to a representation of $C^*_\alpha(\mathbb Z^2)$.

The map $\R^2 \ni p \mapsto \epsilon_p \in \left(\Z^2\right)^\sharp$ given by
$\epsilon_p(x):=e^{2\pi ix\cdot p}\equiv \exp\{2\pi i\sum_{j=1,2} x_jp_j\}$
is a surjective group morphism with kernel $\mathbb Z^2$. It follows that
the dual group $\left(\Z^2\right)^\sharp$ can be identified
with the quotient $\T^2:=\R^2/\Z^2$.
We denote by $\mathfrak{C}^{\omega_\alpha}(\Z^2)$ the envelopping
$C^*$-algebra of $\overline{\mathcal{F}}_{\Z^2}\big(L^1(\Z^2)\big)$ and define
$\mathfrak{Op}^\lambda=\mathfrak{Rep}^\lambda\circ
\overline{\mathcal{F}}^{-1}_{\Z^2}$.
Then for $f\in C(\mathbb{T}^2)$, denoting by $\hat{f}_y$ its Fourier
coefficient in $y$, we have the following representation on $\mathcal{H}$:
$$
\left[\mathfrak{Op}^\lambda (f)u\right](x)=\sum_{y\in\Z^2}\omega_\alpha(x,y)
\hat{f}_y u(x+y)
$$
For the Moyal product we obtain:
$$
(f\circ^{\omega_\alpha} g)(\theta):=\sum_{x\in\Z^2}\sum_{y\in\Z^2}
\int_{\mathbb{T}^2}d\tau\int_{\mathbb{T}^2}d\gamma\; e^{2\pi i\theta\cdot x}\;
e^{-2\pi i\tau\cdot y}\;e^{-2\pi i\gamma\cdot (x-y)}\;\omega_\alpha(y,x)
\;f(\tau)\; g(\gamma)=
$$
$$
=\sum_{x\in\Z^2}\sum_{y\in\Z^2}
e^{2\pi i\theta\cdot x}\;\omega_\alpha(y,x)\hat{f}_y\;\!\hat{g}_{x-y}.
$$

The dependence on $\alpha$ of the almost Mathieu Hamiltonian was hidden;
but of course its spectral properties depend heavily on the $C^*$-algebra
$C^*_\alpha(\mathbb Z^2)$ in which we consider it embedded.
By representing it via $\mathfrak{Rep}^\lambda$ (other representations are
also interesting), the $\alpha$-dependence becomes explicit. The study of
the resulting operators is one of the main trends in modern spectral theory.

We do not have  much to say about cases more complicated than $\A=\mathbb C$,
because we do not understand $2$-cocycles in these situations (in contrast
whith $X=\mathbb R^N$, for which $\A$-smooth magnetic fields define
$2$-cocycles; see Section \ref{s4}). We just note that ``weighted
convolution operators'' with very general weights are within reach.
Let $\lambda:\mathbb Z^N\times\mathbb Z^N\rightarrow\mathbb T$ be such that
$\omega(q;x,y)=\frac{\lambda(q;x)\lambda(q+x;y)}{\lambda(q;x+y)}$ behaves like
a function in $\A$ in the variable $q$, for $x$ and $y$ fixed.
Let $b$ be a complex function on $\mathbb T^N$. Then
\begin{equation*}
[\mathfrak {op}^\lambda(b)u](x)=\sum_{y\in \mathbb Z^N}\lambda(x;y-x)
\hat b_{y-x}u(y)
\end{equation*}
can be given a meaning under various conditions on $b$; see \cite{Nenciu} for
example. This is only a particular case of the pseudodifferential operators
with symbols defined on $\,\mathbb Z^N \times \T^N$ that are covered by our
formalism.
 
\section{The magnetic case}\label{s4}

We shall outline now how the setting and results of Sections \ref{s2} and
\ref{s3} serve in the quantization of a non-relativistic particle without
spin in a variable magnetic field.

We consider a quantum particle without internal structure moving in $X=\R^N$,
in the presence of a variable magnetic field. The {\it magnetic field} is
described by a closed continuous field of 2-forms $B$ defined on $\R^N$.
It is well-known that any such field $B$ may be written as the differential
$dA$ of a field of 1-forms $A$, {\it a vector potential}, that is highly
non-unique (the gauge ambiguity). By using coordinates, one has
$B_{jk}=\partial_jA_k-\partial_kA_j$ for each $j,k=1,\ldots,N$.

In the presence of the field $B=dA$, the prescription (\ref{Weyl}) has to
be modified. This topic was very rarely touched in the literature and
the following wrong solution appears: The minimal coupling principle
says roughly that the momentum $p$ should be replaced with the magnetic
momentum $\pi^A:=p-A$. This originated in Lagrangian Classical Mechanics
and works well also at the quantum level as long as the expressions are
polynomials of order less or equal to $2$. But if one just replaces in
(\ref{Weyl}) $f\left(\frac{x+y}{2},p\right)$ by
$f\left[\frac{x+y}{2},p-A\left(\frac{x+y}{2}\right)\right]$ one gets a
formula which misses the right gauge covariance. Let us denote the result
of this procedure for some function $f$ in phase space by $\Op_A(f)$.
If another vector potential $A'$ is chosen such that $A' = A + \nabla \rho$
with $\rho$ a scalar function, then $dA' = dA$.
But the expected formula $\Op_{A'}(f)=e^{i\rho}\Op_A(f)e^{-i\rho}$
is verified for some simple cases ($A,A'$ linear and $f$ arbitrary, or
$f$ polynomial of order strictly less than 3 in $p$ and $A,A'$ arbitrary),
but in general it fails.

In fact a partial solution was offered long ago in \cite{Luttinger},
in connection with the Peierls substitution, for the case of functions
depending only of $p$ and periodic in this argument. It seems that it
was forgotten, maybe because the most studied case is that of a constant
field, for which the choice of a linear vector potential cannot lead
to any trouble. In \cite{Karasev1}, \cite{Karasev2}, \cite{Purice2}
these matters were tackled systematically,
leading to a new, gauge invariant, pseudodifferential calculus, which was
called {\it the magnetic Weyl calculus}. We review it here as a special,
distinguished case of the preceding developments, underlining its connection
with twisted dynamical systems, cf.~also \cite{Purice1}.

\subsection{Magnetic twisted dynamical systems}\label{s4.1}

From now on $X$ will be the vector space $\R^N$; it is a particular
instance of an abelian, second countable locally compact group.
It is a standard fact that the group dual $X^\sharp$ of $X$ can be
identified with the vector space dual $X^\star$. Let us denote by
$(x,p)\to x\cdot p$ the duality between $X$ and $X^\star$.
Then $\epsilon:X^\star\to X^\sharp$, $\epsilon_p(x):=e^{ix\cdot p}$
is the isomorphism (there are many others but we choose this one).

Particular $2$-cocycles on $X$ will be given by magnetic fields.
{\it A magnetic field on} $X$ is a closed continuous $2$-form $B$.
Since on $X=\mathbb R^N$ we have canonical global coordinates, we shall speak
freely of the components $B_{jk}$ of $B$; they are continuous real
functions on $X$ satisfying $B_{kj}=-B_{jk}$ and $\partial_jB_{kl}+
\partial_lB_{jk}+\partial_kB_{lj}=0$, $\forall j,k,l=1,\dots,N$.
It is well-known that $B=dA$ for some $1$-form $A$ on $X$,
called {\it a vector potential}. The differential is viewed in a
distributional sense. The vector potential is highly non-unique. We restrict only
to continuous $A$; this is always possible, at least by {\it the transversal
gauge}
\begin{equation}\label{trans}
A_j(x):=-\sum_{k=1}^N\int_0^1ds\,B_{jk}(sx)sx_k.
\end{equation}

Given a $k$-form $C$ on $X$ and a compact $k$-surface $\gamma\subset X$,
we define $$\Gamma^C(\gamma):=\int_\gamma C,$$
this integral having a well-defined parametrization independent meaning. We
shall mainly encounter circulations of 1-forms along linear segments
$\gamma=[x,y]$ and fluxes of 2-forms through triangles $\gamma=<x,y,z>$. 

For a continuous magnetic field $B$ one defines
\begin{equation}\label{coma}
\omega^B(q;x,y) := e^{-i\Gamma^B (< q,q+x,q+x+y>)}\qquad
\hbox{for all } x,y,q \in X.
\end{equation}

Let us now fix a separable $X$-algebra $\A$ with spectrum $\SA$. 
Functions of type $\A$ (see Definition \ref{tipA}) are continuous on $X$, but
one observes that they can be unbounded if $\A$ is not unital.
The simplest example is obtained by considering $\A = \CC_0(X)$ with
$\SA$ equal to $X$ and $\delta^\A$ the identity on $X$. In the sequel we
shall supress the notational difference between $b$ and $\tilde b$ from
Definition \ref{tipA}.

\begin{definition}
A magnetic field $B$ is of type $\A$ if all its components
$\{B_{jk}\}_{j,k=1,\dots,N}$ are of type $\A$.
\end{definition}

\begin{lemma}\label{ordine}
If $B$ is a magnetic field of type $\A$, then $(\A,\theta,\omega^B,X)$
is a standard twisted dynamical system.
\end{lemma}

\begin{proof}
The proof that $\omega^B$ is a normalized 2-cocyle, i.e.~it satisfies relations
(\ref{cocond}) and (\ref{condi2}), follows easily by direct calculations (for
the first one use the Stokes Theorem for the closed 2-form $B$ and the
tetrahedron of vertices $q, q+x, q+x+y, q+x+y+z$).

We show that $\omega^B$ has the right continuity properties.
It should define a mapping
\begin{equation}\label{coc-magn}
X\times X\ni(x,y)\to\left[\omega^B(x,y)\right](\cdot)\equiv\omega^B(\cdot;x,y)
\in\CC(\SA;\mathbb T),
\end{equation}
continuous with respect to the topology of uniform convergence on compact
subsets of $\SA$. But this is equivalent to the fact that $\omega^B$
defines un element of $\CC(\SA\times X\times X;\T)$; this type of statement
has already appeared in the proof of Lemma \ref{triv}. Taking into account
obvious properties of the exponential, this amounts to the fact that the
function
$$
\varphi^B:X\times X\times X\rightarrow \mathbb R,\ \ \varphi^B(q;x,y)
:=\Gamma^B(<q,q+x,q+x+y>)
$$
can be viewed (using the map $\delta^\A$ at the level of the first variable)
as a continuous function on $\SA\times X\times X$.

We use the parametrization
\begin{equation*}
\varphi^B(q;x,y)=\sum_{j,k=1}^Nx_j\;\!y_k
\int_0^1dt \int_0^1 ds\;\!s\;\!B_{jk}(q+sx+sty).
\end{equation*}
The continuous action $\theta$ defines a continuous mapping
$\,\tilde\theta:X\times \SA \rightarrow \SA$, so one has the continuous
correspondence $\SA\times X\times X\ni(q;x,y)\to q+sx+sty=
\tilde\theta_{sx+sty}(q)\in \SA$. Since $B_{jk}$ is seen as a continuous
function: $\SA\to\mathbb R$, the assertion follows easily.
\end{proof}

We call $(\A,\theta,\omega^B,X)$ {\it the twisted dynamical system
associated with the abelian algebra $\A$ and the magnetic field} $B$.
In most of the cases the $2$-cocycle $\omega^B\in Z^2\big(X;\U(\A)\big)$
is not trivial. But as Proposition \ref{Couchepin} shows, it is pseudo-trivial.
In fact, its pseudo-trivialization can be achieved by a vector potential.
Any continuous $1$-form $A$ defines a $1$-cochain
$\lambda^A\in C^1\big(X;\CC(X;\T)\big)$ via its circulation:
\begin{equation}\label{maco}
\left[\lambda^A(x)\right](q)\equiv\lambda^A(q;x)=e^{-i\Gamma^A([q,q+x])}
= e^{-ix\cdot \int_0^1 ds \;\!A(q + sx)}.
\end{equation}
As soon as $dA=B$, we have $\delta^1(\lambda^A)=\omega^B$
$\big($a priori with respect to $\CC(X;\T)\big)$,
by a suitable version of Stokes Lemma. As said above, the transversal gauge
offers a continuous vector potential corresponding to a given $B$. Actually,
this is consistent with the choice (\ref{psetri}) of a pseudo-trivialization
of $\omega^B$: for $q,x\in X$, $\lambda(q;x):=\omega^B(0;q,x)=
e^{-i\Gamma^B<0,q,q+x>}$ and it follows immediatly that $\Gamma^B(<0,q,q+x>)
=\Gamma^A([q,q+x])$, with $A$ given by (\ref{trans}).

Depending on the $X$-algebra $\A$, different magnetic fields can give
cohomologous $2$-cocycles:

\begin{definition}
{\rm Let $B_1$, $B_2$ be two magnetic fields of type $\A$. We say that
they are $\A${\it -equivalent} if there exists a vector potential $A$ with
components of type $\A$ such that $B_2-B_1=dA$.}
\end{definition}

\begin{proposition}
If $\,B_1$, $B_2$ are $\A$-equivalent, then $\,\omega^{B_1}$ and
$\,\omega^{B_2}$ are cohomologous.
\end{proposition}

\begin{proof}
If the components of $A$ are of type $\A$, then considerations as in
the proof of Lemma \ref{ordine} show that $\lambda^A\in C^1\big(X;\U(\A)\big)
=C^1\big(X;\CC(\SA;\T)\big)$. In addition $\,\omega^{B_2}=\omega^{B_1}
\delta^1(\lambda^A)$.
\end{proof}

We explained above that any two magnetic fields that are continuous on $X$
$\big($this means exactly that they are of type $\CC_0(X)\big)$ are also
$\CC_0(X)$-equivalent. On the other hand two constant magnetic fields
(of type $\mathbb C$) are $\mathbb C$-equivalent only if they coincide.

Let us point out that the natural approach, in the framework of twisted
dynamical systems associated with magnetic fields, would be to start 
with a given magnetic field and consider the translational invariant 
$C^*$-algebra $\mathcal{A}$ that it generates. The magnetic field will be 
of type $\mathcal{A}$ by construction. 
Then, one may enlarge this $C^*$-algebra in order to fit some other 
asymptotic behaviours associated with the operators one intends
to consider (see also \cite{Georgescu1}, \cite{Purice1}).

Another interesting subject would be to understand to what extent the
2-cocycles and the group cohomology discussed in Subsection 
{\bf \ref{cohomol}} may be related to magnetic fields (as above) and 
with de Rham cohomology. It is evident that requiring the cocycles to 
be multipliers on some given $C^*$-algebra induces important obstructions. 
Moreover, although de Rham cohomology of $\R^N$ is trivial, the vector 
potentials associated with some bounded magnetic field may no longer define 
good multipliers. 

Now, specific standard twisted dynamical systems being constructed, the whole
formalism of the preceding sections is released. {\it The twisted crossed
product} $\A\rtimes^{\omega^B}_{\theta,\tau}X$, denoted slightly simpler
by $\A\rtimes^{B}_{\theta,\tau}X$, {\it is said to be assigned to the
$X$-algebra $\A$, the magnetic field $B$ and the endomorphism $\tau$.}
As always, the dependence on $\tau$ is within isomorphism. If one replaces
$B$ by some $\A$-equivalent $B'$, the $C^*$-algebras
$\A\rtimes^{B}_{\theta,\tau}X$ and $\A\rtimes^{B'}_{\theta,\tau}X$ will be
canonically isomorphic. For any continuous $B$ the $C^*$-algebra
$\CC_0(X)\rtimes^{B}_{\theta,\tau}X$ is isomorphic to $\K(\H)$, the ideal of
all compact operators in $\H=L^2(X)$.

The fact that the magnetic $2$-cocycle $\omega^B$ satisfies
\begin{equation}\label{this}
\omega^B(q;sx,tx)=1,\ \ \forall q,x \in X\ \ \text{and}\ \ \forall s,t\in\mathbb R
\end{equation}
leads directly to the magnetic momenta. Let us fix some continuous $A$ such
that $dA=B$, thus $\delta^1(\lambda^A)=\omega^B$ $\big($pseudo-trivialization with
respect to $\CC(X;\mathbb T)\big)$. Then $\lambda^A$ will satisfy for all
$q,x\in X$ and all $s,t\in\mathbb R$: $\lambda^A(q;sx+tx)=\lambda^A(q;sx)
\lambda^A(q+sx;tx)$ (in general, if $\lambda$ is not the exponential of
{\it a circulation} this will not be true). 
We consider {\it the Schr\"odinger
covariant representation $(\H,r,T^{\lambda^A}\equiv T^A)$ defined by} $A$,
with $\H=L^2(X)$, $r(a)=a(Q)$ and
$$
[T^A(y)u](x)=\lambda^A(x;y)u(x+y),\ \ x,y\in X,\ \ u\in\H.
$$
The unitary operators $\{T^A(y)\}_{y\in X}$ are called {\it magnetic
translations}. They appear often in the physical literature; we refer to
\cite{Luttinger} and \cite{Zak} for example. One has, by a short calculation,
\begin{equation}\label{Stone}
T^A(sx+tx)=T^A(sx)T^A(tx),\ \ \forall x\in X,\ \ \forall s,t\in \mathbb R
\end{equation}
and this also implies $T^A(-x)=T^A(x)^{-1}\left(=T^A(x)^*\right)$,
$\,\forall x\in X$. In fact, the formula
$$
T^A(y)T^A(z)=r[\omega^B(y,z)]T^A(y+z),\ \ y,z\in X
$$
shows that (\ref{Stone}) is equivalent with (\ref{this}). For $t\in\mathbb R$
and $x\in X$, let us set $T^A_t(x):=T^A(tx)$. By (\ref{Stone}),
$\{T^A_t(x)\}_{t\in\mathbb R}$ is an evolution group in $\H$ for any $x$.
Thus, by Stone Theorem, it has a self-adjoint generator that moreover depends
linearly (as a linear operator on $\H$) on the vector $x\in X$; thus
we denote it by $x\cdot\Pi^A$ and call it {\it the projection on $x$ of the
magnetic momentum associated with the vector potential} $A$. For any index
$j\in\{1,...,N\}$ we denote $\Pi^A_j:=e_j\cdot \Pi^A$ the projection of
the magnetic momentum on the $j$'th vector of the canonical base in $X$.
A direct calculation shows that on $C_c^\infty(X)$ one has
$\Pi^A_j=-i\partial_j-A_j(Q)$.

\subsection{Magnetic pseudodifferential operators}

One remote origin of the present work is the observation that a certain
tentative to quantize systems in a variable magnetic field fails,
lacking gauge-covariance.

We recall from {\bf \ref{sub11}} the formula (\ref{Weyl}) giving the Weyl
prescription to quantize a symbol $f$ defined on the phase-space
$\Xi=X\times X^{\star}$. The background is a physical system composed of
a non-relativistic, spinless particle moving in the configurational space $X$
in the absence of any magnetic field. When a magnetic field $B$ is turned on,
a tentative to incorporate it was via the minimal coupling principle.
This states, rather vaguely, that $B$ can be taken into account by choosing a
vector potential $A$ corresponding to $B$ and replacing the canonical
variables $(x,p)$ by $\big(x,p-A(x)\big)$. This would lead to the formula
\begin{equation}\label{prost}
[\Op_A(f)u](x):=\int_{\R^{2N}}dy\;\!dp\;e^{i(x-y)\cdot
p}f\left[\frac{x+y}{2},p-A\left(\frac{x+y}{2}\right)\right]u(y)
\end{equation}
which, in its turn, imposes a certain symbolic calculus, modifying the
formula (\ref{Moyal}). In fact this cannot be correct, since under a change
of the vector potential $A'=A+\nabla\rho$ one fails to obtain the natural
covariance relation $\Op_{A'}(f)=e^{i\rho(Q)}\Op_A(f)e^{-i\rho(Q)}$.

The solution was offered (independently) in \cite{Karasev1} and
\cite{Purice2}. We refer also to \cite{Purice1} and \cite{Karasev2} to some
related works and note that the right attitude (undeveloped and stated only
within a restricted setting) has already appeared in \cite{Luttinger}.
Nevertheless the formula (\ref{prost}) still works in two important particular
cases, which were studied mostly. One is that of a linear vector potential $A$
(giving a constant magnetic field) and an arbitrary (reasonable) symbol $f$.
The other is obtained by taking an arbitrary (continuous) $A$ but restricting
$f$ to be a polynomial of order $\le 2$ in $p$ with constant coefficients.
We refer to \cite{Purice2}, Subsection 3.4 for a discussion on this point.

In the present context, quantizing the
above physical system in the presence of the magnetic field follows as a
particular case of the developments in Sections {\bf \ref{s2}} and {\bf
\ref{s3}}. At the level of twisted dynamical systems and twisted crossed
products the particularization was initiated in {\bf \ref{s4.1}}.
The situation in pseudodifferential terms will be discussed now.
This could serve as a useful r\'esum\'e for the reader. Everything has
already been proved at a more general level.

So let $B$ be a continuous magnetic field and $A$ a corresponding continuous
vector potential. The $2$-cocycle $\omega^B$ and the $1$-cochain $\lambda^A$
are defined, repectively, in (\ref{coc-magn}) and (\ref{maco}).
In our case $X=\mathbb R^N$ and $\tau=1/2$ is an endomorphism of $X$.
It leads to the most symmetric formulae, so we shall restrict to $\tau=1/2$
for simplicity. It was shown before that other choices lead to isomorphic
formalisms.

\paragraph{The magnetic Weyl system.}

For $X=\mathbb{R}^N$, $\Xi:=X\times X^\star$ is the usual phase space with the
symplectic structure defined by the canonical symplectic form
\begin{equation*}
\sigma[(x,p),(y,k)]:=y\cdot p-x\cdot k,\qquad (x,p),(y,k)\in\Xi.
\end{equation*}

Associated with the Schr\"{o}dinger covariant representation
$(\H,r,T^A)$ defined above, we can define now the {\it
  magnetic Weyl system} $W^A$
$$
\Xi\ni (x,p)\mapsto W^A\big((x,p)\big):=e^{-\frac{i}{2}x\cdot p}\;\!
e^{-iQ \cdot p}\;\!T^A(x)\in\mathcal{U}(\H).
$$
These unitary operators define then a projective representation satisfying
$$
W^A(\xi)\;\!W^A(\eta)=e^{\frac{i}{2}\sigma(\xi,\eta)}\;\!\omega^B(x,y)
\;\!W^A(\xi+\eta)
$$ 
where $\xi=(x,p)$, $\eta=(y,k)$.

The symplectic structure comes into play at a rather late stage because at the
beginning we were placed in a very general setting. Anyway, our starting point
was ``a twisted action of the group $X$ on an abelian algebra of position
observables'' and 
this is nearly an extension of the framework of ``imprimitivity systems'',
cf.~\cite{Varadarajan}. But it is perfectly natural (and well-suited to the
classical picture) to start from a symplectic formalism. This is done in
\cite{Karasev1} and \cite{Karasev2}. The canonical symplectic form and the
magnetic field are melt together into a $B$-depending symplectic form and this
is at the basis of the quantization procedure. A beautiful geometric picture
emerges, for which we refer to the quoted papers.

\paragraph{The magnetic Weyl calculus.}

For any $f\in\mathcal{S}(\Xi)$ we define the operator
$\Op^A(f):=\Op^{\lambda^A}_{1/2}(f)$ in $\H=L^2(X)$ and obtain 
$$
\left[\Op^A(f)u\right](x)=\int_X\int_{X^\star}dy\,dp\,e^{i(x-y)\cdot p}
e^{-i\Gamma^A([x,y])}f\left(\frac{x+y}{2},p\right)u(y).
$$
It is easy to observe that this is an integral operator with kernel
$$
K^A:=\tilde{\lambda}^A\;\!S^{-1}\;\!(\mathbf{1}\otimes\FRb)
$$
where $\tilde{\lambda}^A(x,y):=\lambda^A(x;y-x)$ and $\big(S^{-1}h\big)
(x,y)=h\left(\frac{x+y}{2},x-y\right)$. It is now
easy to extend the map $K^A$ and thus define $\Op^A(F)$ for any
$F\in\mathcal{S}^\prime(\Xi)$ as the integral operator with kernel $K^A(F)$,
defined on $\mathcal{S}(X)$ with values in $\mathcal{S}^\prime(X)$.
It seems legitimate to view the correspondence $f\to\Op^A(f)$ as a
functional calculus for the family of self-adjoint operators
$Q_1,\dots,Q_N,\Pi^A_1,\dots,\Pi^A_N$. The high degree of non-commutativity
of these $2N$ operators stays at the origin of the sophistication of the
symbolic calculus. The commutation relations
\begin{equation}\label{macom}
i[Q_j,Q_k]=0,\ \ i[\Pi^A_j,Q_k]=\delta_{jk},\ \ i[\Pi^A_j,\Pi^A_k]=-B_{jk}(Q),
\ \ \ j,k=1,\dots,N
\end{equation}
collapse for $B=0$ to the canonical commutation relations satisfied by $Q$
and $P$. But they are much more complicated, especially when $B$ is not a
polynomial. In particular, no simple analogue of the Heisenberg group
is available. The main mathematical miracle that allows, however, a nice
treatment is the fact that (\ref{macom}) can be recast in the form of a
covariant representation of a twisted dynamical system. And this is connected
to its symplectic flavour.

Let us emphasis here that the functional calculus that we have defined is
gauge covariant, in the sense that it satisfies the property: 
If $A'=A+\nabla\varphi$ with $\varphi:X\rightarrow\mathbb R$ continuous,
then $\Op^{A'}(f)=e^{i\varphi(Q)}\Op^A(f)e^{-i\varphi(Q)}$. This gauge
covariance property may be seen as a special instance of
Proposition \ref{formula} (b).

\paragraph{The magnetic symbolic calculus.}

The usual product and adjoint operation on $\mathcal{B}(\mathcal{H})$ allows
now to define on $\mathcal{S}(\Xi)$ a composition and an involution:
\begin{equation}\label{leg}
(f\circ^B g)(\xi)=4^N\int_\Xi\int_\Xi
d\eta\,d\zeta\,e^{-2i\sigma(\xi-\eta,\xi-\zeta)}
e^{-i\Gamma^B(<q-y+x,x-q+y,y-x+q>)}f(\eta)g(\zeta),
\end{equation}
$\big($for $\xi = (q,l)$, $\eta = (x,p)$ and $\zeta = (y,k)\big)$
$$
f^{\circ^B}(\xi)
=\overline{f(\xi)},\quad\forall \xi\in\Xi
$$
such that
\begin{equation*}
\Op^A(f\circ^B g)=\Op^A(f)\Op^A(g),\ \ \Op^A(f^{\circ^B})=\Op^A(f)^*,
\ \ (f\circ^B g)^{\circ^B}=g^{\circ^B}\circ^B f^{\circ^B}.
\end{equation*}

Let us remark that the involution $^{\circ^B}$ and the product $\circ^B$ are
defined intrinsically, without any choice of a vector potential. The choice is
only needed when we represent the resulting structures in a Hilbert space. We
call (\ref{leg}) {\it the magnetic Moyal product}. 
The involution $^{\circ^B}$ does not depend on $B$ at all. This is no longer true
if $\tau\ne 1/2$. The property $\omega^B(x,-x)=1$, $\forall x\in X$ is also
used to get the simple form of $^{\circ^B}$.

Let us assume now that $B$ is of type $\A$ for some $X$-algebra $\A$.
The $C^*$-algebra $\mathfrak C^{\omega^B}_{\A,1/2}$, introduced in
{\bf \ref{s3.1}}, will be denoted by $\mathfrak C^B_\A$. We call it
{\it the $C^*$-algebra of pseudodifferential symbols of class $\A$
associated with $B$}. We recall that it is essentially a partial Fourier
transform of the twisted crossed product $\A\rtimes ^B_{\theta,1/2} X$.
The formulae defining the magnetic Weyl calculus make sense at least on
the dense subset $\mathfrak F L^1(X;\A)$, with iterated integrals.
The extension of $\Op^A$ is a faithful representation of the $C^*$-algebra
$\mathfrak C^B_\A$ for any continuous $A$ with $dA=B$. If $\CC_0(X)\subset\A$,
then $\Op^A$ is irreducible.

\begin{remark}
{\rm (1) If $B$ and $B'$ are of type $\A$ and $\A$-equivalent, then the
$C^*$-algebras $\mathfrak C^B_\A$ and $\mathfrak C^{B'}_\A$ are isomorphic.\\
(2) In particular, $\mathfrak C^B_{\CC_0(X)}$ and $\mathfrak
C^{B'}_{\CC_0(X)}$ are always isomorphic if $B$ and $B'$ are of type
$\CC_0(X)$ (and this means just that the components $B_{jk}$, $B'_{jk}$
are continuous functions on $X$, maybe unbounded). In this case, for any
continuous $A$ defining $B$, $\Op^A$ sends $\mathfrak C^B_{\CC_0(X)}$
isomorphically on $\K[L^2(X)]$.}
\end{remark}

It is very useful that the usual pseudodifferential calculus can be
extended to unbounded symbols. This works well for the magnetic version.
The modest extension strategy of {\bf \ref{s3.3}} can now be greatly
improved; in $X=\mathbb R^N$ there are plenty of extra structures that
can be exploited. First of all, one notes that $\Op^A$ and $\Rep^A$ are
integral operators with $L^2$ kernels. But if the magnetic field is of
class $\CC^\infty_{\text{pol}}$ (i.e. it is $\CC^\infty$ and all its
derivatives verify polynomial bounds) then one can choose $A$ in the same
class and the kernel will have an improved behaviour.
It follows (see \cite{Purice2}, Subsection 3.2) that one can extend,
in a suitable weak sense, $\Op^A$ to temperate distributions;
the same will be true for $\Rep^A$.

It is more difficult to extend significantly the product.
Usually, in a pseudodifferential setting, one tries oscillatory integral
techniques. This gives refined results but the details are sometimes
cumbersome. In \cite{Gracia1} and \cite{Gracia2} it is shown how to extend
the standard Moyal product (\ref{Moyal}) by duality methods, such to obtain
large $^*$-algebras of distributions. Under the assumption that $B$ is in
$\CC^\infty_{\text{pol}}$, this can also be done for the more complicated
magnetic Moyal product (\ref{leg}). This is the subject of Section 4 of
\cite{Purice2}. By a partial Fourier transform or by an analogous direct
approach, this works also for the composition law
$\diamond^B\equiv\diamond^{\omega^B}_{1/2}$. We do not give details here.

{\bf Acknowledgements:} We thank the University of Geneva, where a large part
of  this work was done and especially 
Werner Amrein for his kind hospitality. We also acknowledge partial support
from the CERES Programme (contract no. 38/2002 and no. 28/2003) and the CNCSIS
grant no. 33536/2003.


\begin{thebibliography}{999}

\bibitem[ABG]{Amrein} W.O. Amrein, A. Boutet de Monvel and V. Georgescu:
{\it $\CO$-Groups, Commutator Methods and Spectral Theory of N-Body
Hamiltonians}, Birkh\"auser Verlag, 1996.

\bibitem[AMP]{Amrein'} W.O. Amrein, M. M\u antoiu and R. Purice:
{\it Propagation Properties for Schr\"odinger Operators Affiliated to
Certain $C^*$-Algebras}, Ann. Henri Poincar\'e {\bf 3}, 1215-1232, 2002.

\bibitem[An]{Anderson} R.F. Anderson: {\it The Weyl Functional Calculus},
J. Funct. Anal. {\bf 4}, 240-267, 1969.

\bibitem[AHS]{Avron} J. Avron, I. Herbst and B. Simon: {\it Schr\"odinger
Operators with Magnetic Fields. I. General Theory}, Duke Math. J. {\bf
45}, 847-883, 1978.

\bibitem[Be1]{Bellissard1} J. Bellissard: {\it Non-Commutative Methods
in Semiclassical Analysis, Transition to Chaos in Classical and Quantum
Mechanics} (Montecatini Terme, 1991), 1--64, Lecture Notes in Math.
{\bf 1589}, Springer, Berlin, 1994.

\bibitem[Be2]{Bellissard2} J. Bellissard: {\it Le papillon de Hofstadter}
(d'apr\`es B. Helffer et J. Sj\"ostrand),  S\'eminaire Bourbaki, Vol.
1991/92.  Ast\'erisque {\bf 206}, Exp. No. 745, 3, 7--39, 1992.

\bibitem[Bo] {Boca} F.P. Boca: {\it Rotation Algebras and Almost
Mathieu Operators}, Theta Series in Advanced Mathematics, Theta Foundation,
2001.

\bibitem[BS]{Busby} R. Busby and H. Smith: {\it Representations of Twisted
Group Algebras}, Trans. Amer. Math. Soc. {\bf 149}, 503-537, 1970.

\bibitem[DKR]{Doplicher} S. Doplicher, D. Kastler and D.W. Robinson:
{\it Covariance Algebras in Field Theory and Statistical Mechanics}, Commun.
Math. Phys. {\bf 3}, 1-28, 1966.

\bibitem[FD]{Fell} J.M.G. Fell and R.S. Doran: {\it Representations of
${}^*$-Algebras, Locally Compact Groups, and Banach ${}^*$-Algebraic
Bundles}, Volumes 125 and 126 in Pure and Applied Mathematics,
Academic Press, 1988.

\bibitem[Fo]{Folland} G.B. Folland: {\it Harmonic Analysis in Phase Space},
Princeton University Press, Princeton, New Jersey, 1989.

\bibitem[GI1]{Georgescu1} V. Georgescu and A. Iftimovici: {\it Crossed
Products of $C^*$-Algebras and Spectral Analysis of Quantum
Hamiltonians}, Commun. Math. Phys. {\bf 228}, 519-560, 2002.

\bibitem[GI2]{Georgescu2} V. Georgescu and A. Iftimovici: {\it $C^*$-Algebras
of Energy Observables: II. Graded Symplectic Algebras and Magnetic
Hamiltonians}, Preprint 01-99 at www.ma.utexas.edu/mp\_arc/.

\bibitem[GBV1]{Gracia1} J.M. Gracia-Bond\'ia and J.C. V\'arilly: {\it Algebra
of Distributions Suitable for Phase-Space Quantum Mechanics. I}, J.
Math. Phys. {\bf 29} (4), 869-879, 1988.

\bibitem[GBV2]{Gracia2} J.M. Gracia-Bond\'ia and J.C. V\'arilly:
{\it Algebra of Distributions Suitable for Phase-Space Quantum Mechanics. II.
Topologies on the Moyal Algebra}, J. Math. Phys. {\bf 29} (4), 880-887, 1988.

\bibitem[Gui]{Guichardet} A. Guichardet: {\it Cohomologie des groupes
topologiques et des alg\`ebres de Lie}, Textes Math\'ematiques, 2. CEDIC,
Paris, 1980.

\bibitem[Hor]{Hormander} L. H\"ormander: {\it The Weyl Calculus of
Pseudo-differential Operators}, Comm. Pure Appl. Math. {\bf 32}, 359-443,
1979.

\bibitem[KO1]{Karasev1} M.V. Karasev and T.A. Osborn: {\it Symplectic Areas,
Quantization and Dynamics in Electromagnetic Fields}, J. Math. Phys. {\bf
43} (2), 756-788, 2002.

\bibitem[KO2]{Karasev2} M.V. Karasev and T.A. Osborn: {\it Quantum Magnetic
Algebra and Magnetic Curvature}, Preprint quant-ph/0311053 at xxx.lanl.gov.

\bibitem[Kl1]{Kleppner1} A. Kleppner: {\it Multipliers on Abelian Groups},
Math. Ann. {\bf 158}, 11-34, 1965.

\bibitem[La]{Landsman} K. Landsman: {\it Mathematical Topics Between Classical
and Quantum Mechanics}, Springer-Verlag New York Berlin Heidelberg, 1998.

\bibitem[Le]{Leinfelder} H. Leinfelder: {\it Gauge Invariance of Schr\"odinger
Operators and Related Spectral Properties}, J. Operator Th. {\bf 9},
163-179, 1983.

\bibitem[Lu]{Luttinger} J.M. Luttinger: {\it The Effect of the Magnetic Field
on Electrons in a Periodic Potentials}, Phys. Rev., {\bf 84}, 814-817, 1951.

\bibitem[Ma1]{Mantoiu1} M. M\u antoiu: {\it $C^*$-Algebras, Dynamical Systems
at Infinity and the Essential Spectrum of Generalized Schr\"odinger
Operators}, J. reine angew. Math. {\bf 550}, 211-229, 2002.

\bibitem[Ma2]{Mantoiu2} M. M\u antoiu: {\it $C^*$-Algebras and Topological
Dynamics in Spectral Analysis}, in Operator Algebras and Mathematical Physics,
ed. J.-M. Combes, J. Cuntz, G.A. Elliott, G. Nenciu, H. Siedentop and S.
Str\u atil\u a, Conference Proceedings, Constanta, July 2-7, 2001, 299-314,
2003.

\bibitem[MP1]{Purice1} M. M\u antoiu and R. Purice: {\it The Algebra of
Observables in a Magnetic Field}, Mathematical Results in Quantum
Mechanics (Taxco, 2001), Contemporary Mathematics {\bf 307}, Amer. Math.
Soc., Providence, RI, 239-245, 2002.

\bibitem[MP2] {Purice2} M. M\u antoiu and R. Purice: {\it The Magnetic
Weyl Calculus}, to appear in J. Math. Phys.

\bibitem[Mo]{Moore} C.C. Moore: {\it Group Extensions and Cohomology
for Locally Compact Groups. III}, Trans. Amer. Math. Soc. {\bf 221},
No 1, 1976.

\bibitem[Ne]{Nenciu} G. Nenciu: {\it On the Smoothness of Gap Boundaries
for Generalized Harper Operators}, preprint 03-399 at
www.ma.utexas.edu/mp\_arc/.

\bibitem[Pa]{Packer} J.A. Packer: {\it Transformation Group $C^*$-Algebras:
A Selective Survey}, $C\sp *$-Algebras: 1943--1993 (San Antonio, TX, 1993),
Contemp. Math. {\bf 167}, Amer. Math. Soc., Providence, RI, 182--217, 1994.

\bibitem[PR1]{Raeburn1} J.A. Packer and I. Raeburn: {\it Twisted Crossed
Products of $C^*$-Algebras}, Math. Proc. Camb. Phyl. Soc. {\bf 106},
293-311, 1989.

\bibitem[PR2]{Raeburn2} J.A. Packer and I. Raeburn: {\it Twisted
Crossed Products of $C^*$-Algebras II}, Math. Ann. {\bf 287}, 595-612,
1990.

\bibitem[Pe]{Pedersen} G. Pedersen: {\it $C^*$-Algebras and their Automorphism
Groups}, Academic Press, 1979.

\bibitem[RW]{Williams} I. Raeburn and D.P. Williams: {\it Morita
Equivalence and Continuous-Trace $C^*$-algebras}, Mathematical Surveys
and Monographs, Volume 60, American Mathematical Society, 1998.

\bibitem[Ri1]{Rieffel1} M.A. Rieffel: {\it $C^*$-Algebras Associated with
Irrational Rotations}, Pac. J. Math. {\bf 95} (2), 415-419, 1981.

\bibitem[Ri2]{Rieffel2} M.A. Rieffel: {\it Projective Modules over
Higher-Dimensional Non-Commutative Tori}, Can. J. Math, {\bf 40} (2), 257-338,
1988.

\bibitem[Ri3]{Rieffel3} M.A. Rieffel: {\it Quantization and $C^*$-Algebras},
$C\sp *$-Algebras: 1943--1993 (San Antonio, TX, 1993),
Contemp. Math. {\bf 167}, Amer. Math. Soc., Providence, RI, 182--217, 1994.

\bibitem[Sh1]{Shubin1} M. Shubin: {\it Pseudodifferential Operators and
Spectral Theory}, Springer Series in Soviet Mathematics, Springer-Verlag,
Berlin, 1987.

\bibitem[Sh2]{Shubin2} M. Shubin: {\it Discrete Magnetic Laplacian}, Commun.
Math. Phys. {\bf 164}, 259-275, 1994.

\bibitem[Ta]{Tak} M. Takesaki: {\it Theory of Operator Algebras I},
Springer-Verlag, New York, 1979.

\bibitem[Va]{Varadarajan} V.S. Varadarajan: {\it Geometry of Quantum
Theory, II}, Van Nostrand, Princeton, N.J. 1970.

\bibitem[Za]{Zak} J. Zak: {\it Magnetic Translation Groups}, Phys. Rev. A
{\bf 134}, 1602-1607 and 1607-1611, 1965.

\end{thebibliography}
\end{document}